\documentclass[prb,aps,twocolumn,footinbib,superscriptaddress,floatfix,bibliography]{revtex4-1}

\usepackage{CJK}
\usepackage[colorlinks,bookmarks=false,citecolor=blue,linkcolor=red,urlcolor=blue]{hyperref}
\usepackage{color} 
\usepackage{graphicx}
\usepackage{float}
\usepackage{multirow}
\usepackage{amsmath}
\usepackage{bm}
\usepackage{amsfonts}
\usepackage{braket}
\usepackage{subfigure}
\usepackage{cleveref}

\begin{document}
\title{Exactly solvable spin-$1/2$ XYZ models with highly-degenerate, partially ordered, ground states}

\author{Grgur Palle}
\affiliation{Max Planck Institute for the Physics of Complex Systems, N{\"o}thnitzer Str.~38, Dresden 01187, Germany}
\affiliation{Institute for Theoretical Condensed Matter Physics, Karlsruhe Institute of Technology, Wolfgang-Gaede-Stra{\ss}e~1, Karlsruhe 76131, Germany}
\author{Owen Benton}
\affiliation{Max Planck Institute for the Physics of Complex Systems, N{\"o}thnitzer Str.~38, Dresden 01187, Germany}

\begin{abstract}
Exactly solvable models play a special role in 
Condensed Matter physics, serving as secure theoretical
starting points for investigation of new phenomena.
Changlani \textit{et al}.~[\href{https://journals.aps.org/prl/abstract/10.1103/PhysRevLett.120.117202}{Phys.~Rev.~Lett.~\textbf{120}, 117202 (2018)}] have discovered a limit of the XXZ model for $S=1/2$ spins on the kagome lattice, which is not only exactly solvable, but features a huge degeneracy of exact ground states corresponding to solutions of a three-coloring problem.
This special point of the model was proposed as a parent for multiple phases in the wider phase diagram, including quantum spin liquids.
Here, we show that the construction of Changlani \textit{et al}.\ can be extended to more general forms of anisotropic exchange interaction, finding a line of parameter space in an XYZ model
which maintains both the macroscopic degeneracy and the three-coloring structure of solutions.
We show that the ground states along this line are partially ordered, in the sense that infinite-range correlations of some spin components coexist with a macroscopic number of undetermined degrees of freedom.
We therefore propose the exactly solvable limit of the XYZ model on corner-sharing triangle-based lattices as a tractable starting point for discovery of quantum spin systems which mix ordered and spin liquid-like properties.
\end{abstract}

\maketitle

\section{Introduction}
\label{sec:intro}

Calculations in condensed matter theory must generally bridge
two different gaps. The first is the gap between model and
experiment: any model simple enough to be successfully studied cannot capture every aspect of a real many body system, though we hope to capture the most important and interesting ones. 
The second is the gap between models and the actual calculation of physical quantities.
Even when the model in question
is simple to write down, it is usually necessary to employ some kind of approximation scheme
during calculations.

Sometimes, however, the second gap is absent. 
There are models which are both physically relevant and for which exact calculations are possible. 
These have been important in the development of modern Condensed Matter physics, especially where they have been used to establish the theoretical possibility of novel phenomena or phases of matter.
Notable examples of this include the Shastry-Sutherland 
model~\cite{shastry81}, which established a featureless, gapped ground state in a many-body quantum spin model, and later found realization in SrCu$_2$(BO$_3$)$_2$~\cite{miyahara99, kageyama99}
and the Kitaev honeycomb model~\cite{kitaev06-AnnPhys321},
which a gave an example of a $Z_2$ quantum spin liquid with emergent Majorana fermions, and was later found to be relevant to various spin-orbit coupled magnets~\cite{jackeli09-PRL102, banerjee16, banerjee17, kitagawa18, takagi19}.

\begin{figure}
    \centering
    \includegraphics[width=0.7\columnwidth]{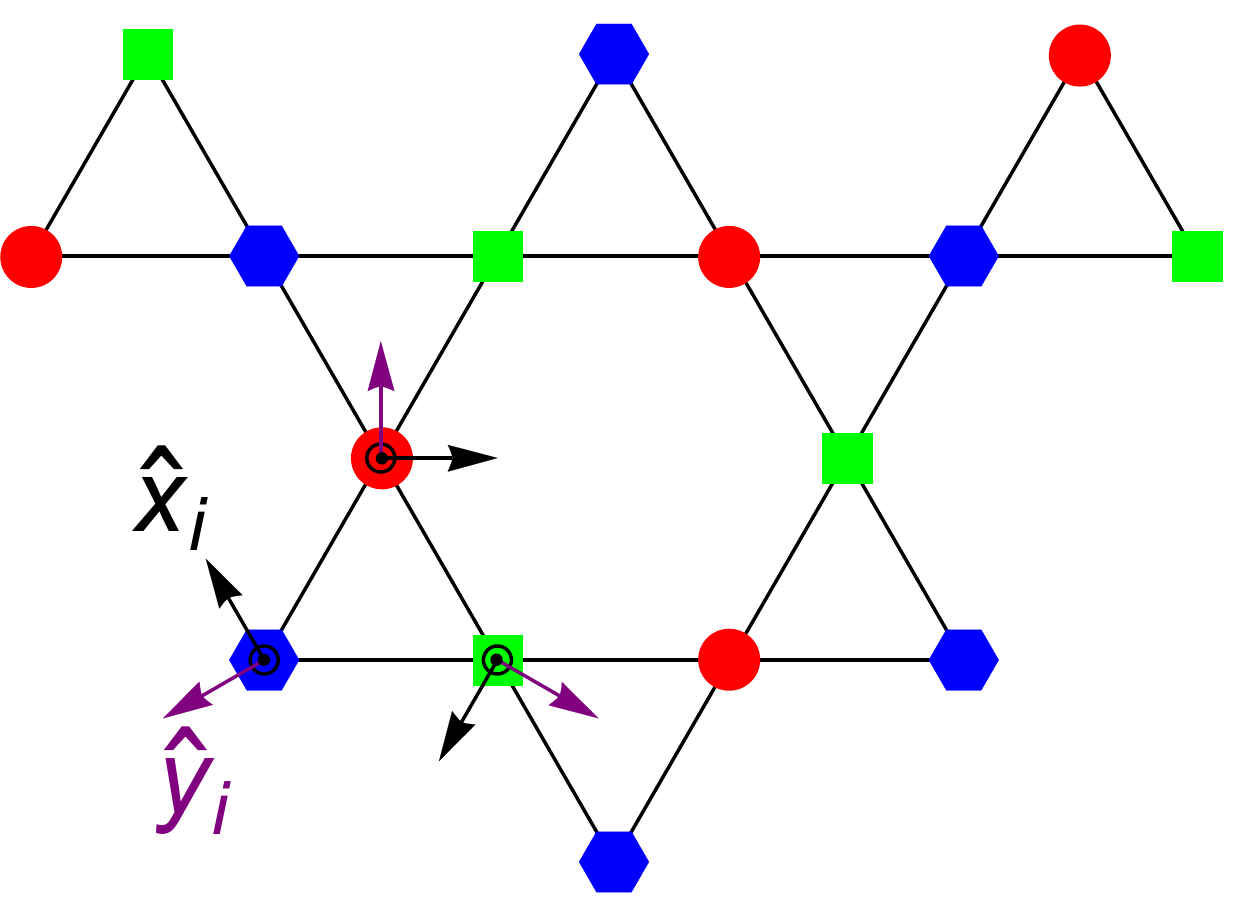}
    \caption{A three-color covering of the kagome lattice. Every triangle must include a site with each of three colors. The number
    of configurations satisfying this local rule grows exponentially with the system size. The exact ground states of the kagome XYZ model [Eq.~(\ref{eq:XYZmodel})] map on to this three-coloring problem when the exchange parameters are chosen to obey Eq.~(\ref{eq:exact-constraint}).
    The coordinate axes $\hat{\bf x}_i$,  $\hat{\bf y}_i$ indicate
    a choice of local basis for the spins, such that Eq.~(\ref{eq:XYZmodel}) is consistent with the symmetry of the kagome lattice, with the $\hat{\bf z}$ axis being uniformly perpendicular
    to the kagome plane. In contrast to the depicted
    coloring, this local spin basis is
    uniform across the lattice.
    }
    \label{fig:kagome_3colors}
\end{figure}

An interesting example of an exactly solvable
spin-1/2 model has been pointed out by
Changlani \textit{et al}.~\cite{changlani18}. They considered
the nearest neighbor XXZ model on the kagome lattice
[Fig.~\ref{fig:kagome_3colors}]
\begin{equation}
\mathcal{H}_{\text{XXZ}} = \sum_{\langle ij \rangle}
\left[
J_{\perp} \left( {S}^x_i {S}^x_j+ {S}^y_i {S}^y_j
\right) + J_z {S}^z_i {S}^z_j
\right]\,,
\label{eq:XXZ}
\end{equation}
and showed that at the special point $J_z/J_{\perp}=-1/2$ (denoted as the XXZ0 point~\cite{Essafi2016})
the model has a set of
exact, degenerate, ground states, the number of which
grows exponentially with system size.
These ground states can be written as simple
product states, despite the fact that the
Hamiltonian is composed of non-commuting terms and that general excited eigenstates are entangled.
The ground states correspond with the solutions of the
three-coloring problem on the kagome lattice~\cite{baxter70}, in which
the vertices of the lattice are colored with three different colors such that no triangle 
has two vertices of the same color
[Fig.~\ref{fig:kagome_3colors}].
For a non-trivial, quantum many-body model to have
such a huge degeneracy of ground states, with such
a simple structure, is remarkable.
The XXZ0 point is also 
significant in the wider phase diagram, being a
point at which several different phases, including distinct quantum spin liquids, meet.
It was suggested that the phase diagram of the model
in the surrounding parameter space could be 
understood starting
from this point~\cite{changlani18, changlani20}.

Here we show how the XXZ0 point
can be generalized to a wider set of
exactly solvable anisotropic $S=1/2$ exchange models.
Specifically, we consider the nearest-neighbor 
XYZ model on the kagome lattice: 
\begin{equation}
\mathcal{H}_{\text{XYZ}}=\sum_{\langle ij \rangle}
\left[
J_x {S}^x_i {S}^x_j+J_y {S}^y_i {S}^y_j+J_z {S}^z_i {S}^z_j
\right]\,,
\label{eq:XYZmodel}
\end{equation}
i.e., the case where each component of the spin has
a different associated exchange constant.
If the basis for the spin operators
$S^{\alpha}_i$ is defined with a local
coordinate frame, as shown in Fig.~\ref{fig:kagome_3colors}, then this model is consistent with the symmetry
of the lattice and can be considered as
a limit of a more general model of anisotropic exchange~\cite{Essafi2017}, as shown in Appendix~\ref{app:gen_models}.

We show that for any parameter set $\{ J \}=( J_x, J_y, J_z )$ fulfilling the conditions:
\begin{equation}
J_x + J_y + J_z > 0
\ \ \text{and} \ \ 
J_x J_y + J_y J_z + J_z J_x =0\,,
\label{eq:exact-constraint}
\end{equation}
there exists a large manifold of exact product ground states,
which generally differs from the solutions of the XXZ0 model, but retains the correspondence with the
three-coloring problem.

We further show that 
despite having an extensive number of undetermined
degrees of freedom,
the ground states
possess infinite range correlations of some spin components, coexisting with algebraic correlations.
This is qualitatively reminiscent of the phenomenon of
``magnetic moment fragmentation''~\cite{brooks14, rougemaille19, lhotel20} studied in both
kagome~\cite{paddison16, canals16, dun20} and
pyrochlore~\cite{brooks14, petit16, benton16, lefrancois17} systems, but is particularly
remarkable because it occurs in the exact ground states of a quantum model.
This is in contrast to most known examples of moment
fragmentation, which either occur in settings where the spins
can be considered classical~\cite{brooks14, paddison16, canals16, lefrancois17}
or as a feature
of the semi-classical dynamics rather than of the ground state~\cite{petit16, benton16}.

The Article is structured as follows:
Section~\ref{sec:model} shows the construction of the family of exactly solvable models and their highly-degenerate ground states;
Section~\ref{sec:order} demonstrates
the partial order of these ground states;
Section~\ref{sec:excitations} demonstrates the presence of gapless excitations above these ground states despite the absence of continuous symmetry in the Hamiltonian; 
Section~\ref{sec:recipe} discusses how our construction can be generalized to other lattices and spin lengths and gives a criterion for the presence of exact ground states of the type discussed here;
lastly, Section~\ref{sec:summary} contains a brief summary of our results and outlook for future work.

\section{Construction of exactly solvable Hamiltonian and its ground states}
\label{sec:model}

In this Section we show how to obtain the result
that the ground states of
Eq.~(\ref{eq:XYZmodel}) can be found  
exactly for all sets of parameters satisfying~(\ref{eq:exact-constraint}).

Any nearest-neighbor Hamiltonian on the
kagome lattice can be written as a sum
of single-triangle Hamiltonians:
\begin{equation}
\mathcal{H}_{\text{XYZ}}=\sum_{\triangle} 
\mathcal{H}_{\text{XYZ}, \triangle}\,.
\end{equation}

The spectrum of the single triangle Hamiltonian $\mathcal{H}_{\text{XYZ}, \triangle}$ 
is composed of a quadruplet with energy
\begin{equation}
e_{q}=-\frac{1}{4} (J_x+J_y+J_z)\,,
\end{equation}
and of two doublets with energies
\begin{equation}
e_{d \pm}= - e_{q} \pm 2 \sqrt{e_{q}^2 - \frac{3}{16} \Lambda}\,\,,
\end{equation}
where 
$
\Lambda=J_x J_y+J_y J_z+J_z J_x\,.
$
When $\Lambda=0$ and $e_q<0$ (cf.~Eq.~(\ref{eq:exact-constraint})), the spectrum is
composed of a 6-fold degenerate ground state $e_q=e_{d-}=e_0$, and of an excited doublet $e_{d+}$.

In this case it follows that the Hamiltonian can
be written as a sum of non-commuting projectors:
\begin{equation}
\mathcal{H}_{\text{XYZ}, \Lambda=0}=\sum_{\triangle}
[(e_{d +}-e_{0}) P_{\triangle} + e_{0}]\,,
\label{eq:projH}
\end{equation}
with $P_{\triangle}$ being the operator
that projects onto the pair of single-triangle
excited states:
\begin{align}
P_{\triangle} &= \lvert d+, + \rangle\langle d+, + \rvert \ \ + \ \ \lvert d+, -\rangle\langle d+, - \rvert \,, \label{eq:tri_ex}
\\
\lvert d+, +\rangle &= \cos(\alpha) \frac{\lvert \uparrow \uparrow
\downarrow \rangle + \lvert \uparrow \downarrow
 \uparrow \rangle + \lvert \downarrow \uparrow \uparrow \rangle}{\sqrt{3}} + \sin(\alpha) \lvert  \downarrow \downarrow \downarrow  \rangle \,, \nonumber \\
\lvert d+, -\rangle &= \cos(\alpha) \frac{\lvert \downarrow \downarrow
\uparrow \rangle + \lvert \downarrow \uparrow
 \downarrow \rangle + \lvert  \uparrow \downarrow \downarrow \rangle}{\sqrt{3}} +
 \sin(\alpha) \lvert  \uparrow \uparrow \uparrow  \rangle \,, \nonumber\\
\tan(\alpha)&=\frac{(J_x-J_y)}{\sqrt{3}(J_x+J_y)}\,. \nonumber
\end{align}

Since $e_{d+}>e_0$, the coefficient in front of the projection operator
in Eq.~(\ref{eq:projH}) is positive and any state which is annihilated
by all of the $P_{\triangle}$ is a ground state. 

We can search for states annihilated by $P_{\triangle}$ amongst the set of site-product
states:
\begin{eqnarray}
\hspace{-1cm} \lvert \Psi (\{\theta\}, \{\phi\}) \rangle= \prod_{{\rm sites} \ j} 
\bigg(&& \cos\left( \tfrac{1}{2} \theta_j \right) e^{-i \tfrac{1}{2} \phi_j} \lvert \uparrow_j \rangle 
\nonumber \\[-8pt]
&& ~~ + \sin\left( \tfrac{1}{2} \theta_j \right) e^{i \tfrac{1}{2} \phi_j}
\lvert \downarrow_j \rangle \bigg)\,. \label{eq:product-wf}
\end{eqnarray}
As we show below, there are many 
product states on the lattice which 
are annihilated by $P_{\triangle}$.
Our strategy is to first identify these
product states on a single triangle (Section~\ref{subsec:product_gs})
and then generate ground states 
on the lattice by tiling single triangle ground states across the system.

\subsection{Product ground states on a single triangle}
\label{subsec:product_gs}

For a single triangle, the possible product wave functions [Eq.~(\ref{eq:product-wf})] are parametrized by $6$ variables:
the $\theta_j$ and $\phi_j$ of each of the three sites.
To be ground states, these wavefunctions need to satisfy
four constraints: the real and imaginary
parts of their overlaps with $\lvert d+, \pm\rangle$ [Eq.~(\ref{eq:tri_ex})]
must vanish.
This suggests a $6-4=2$ dimensional surface of exact
product state ground states for a single
triangle. We have verified that
such a surface, that we shall call $\mathcal{M}$,
indeed exists for all $\{ J \}$ obeying~(\ref{eq:exact-constraint}).
An exact parameterization of $\mathcal{M}$ has also been
found, but its derivation is quite involved and is therefore
presented in Appendix~\ref{app:derivation}.

\begin{figure}[t]
    \centering
    \includegraphics[width=\columnwidth]{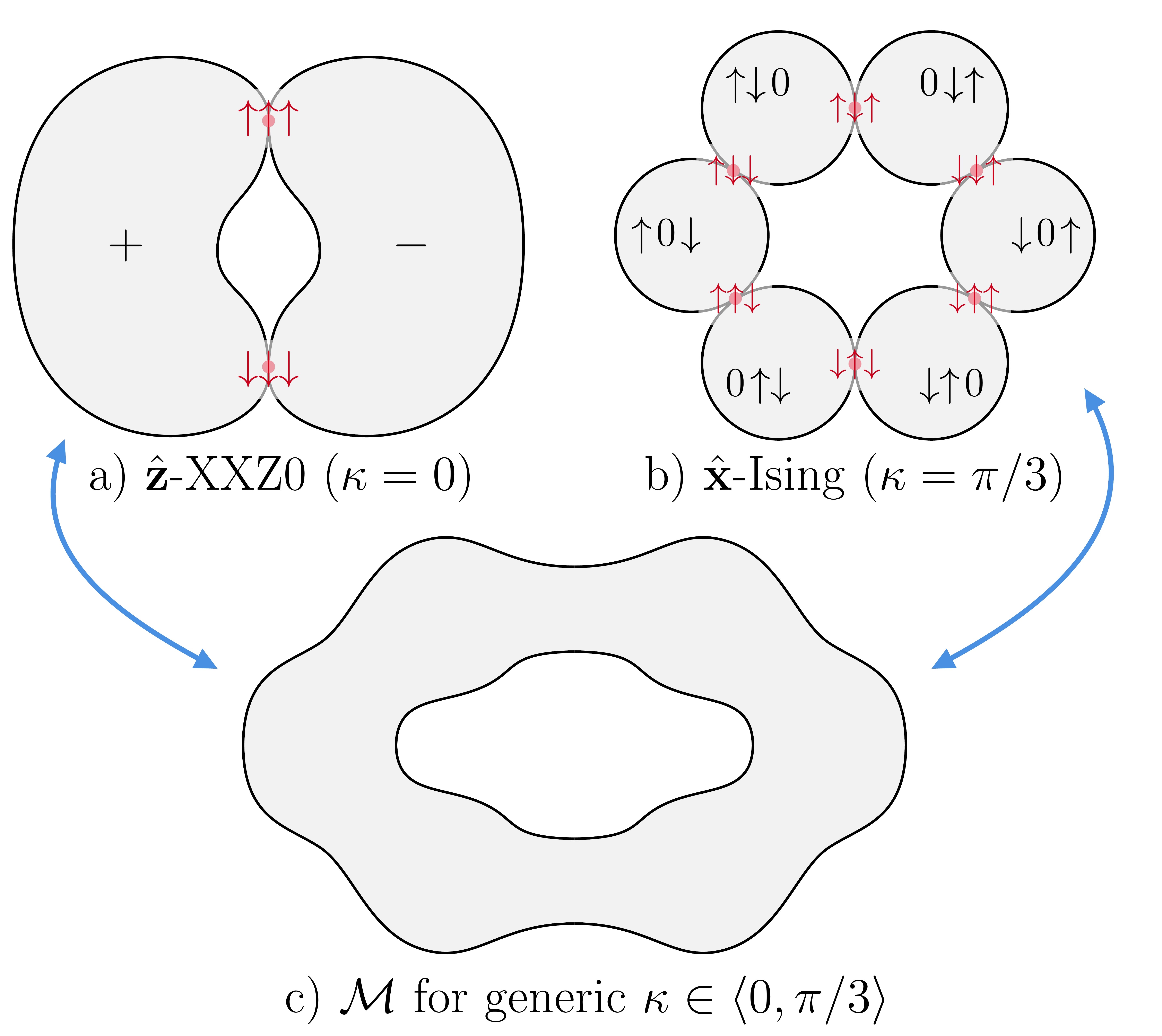}
    \caption{
    Schematic representation of the topology of the
    product ground state manifold of a triangle $\mathcal{M}$ for:
    a) the $\hat{\bf z}$-XXZ0 point ($\kappa=0$, $J_x=J_y=2/3$, $J_z=-1/3$),
    b) the $\hat{\bf x}$-Ising point ($\kappa=\pi/3$, $J_x=1$, $J_y=J_z=0$), and
    c) generic $\kappa \in \langle 0,\pi/3 \rangle$ between these two limits.
    The $\mathcal{M}$ of the XXZ0 point is made of two spheres
    with different chiralities, denoted $+$ and $-$, that touch
    at their poles.
    The points of contact are marked with red dots,
    and the respective spin configurations are written on top
    of these contact points.
    The Ising point $\mathcal{M}$ is made of six spheres that
    have one $\uparrow$ spin, one $\downarrow$ spin, and one free spin,
    which we denoted with a $0$. These six spheres touch adjacent
    spheres at the specified red points.
    In between, $\mathcal{M}$ has the topology of a torus.
    Although drawn here as (deformed) circles, lines, etc., all of the above
    sketches should be understood as representing 2D surfaces
    embedded in the 6D space of all possible $\theta_j, \phi_j$
    in a triangle.}
    \label{fig:special_point_topology}
\end{figure}

Nonetheless, the solutions for
two limits of~(\ref{eq:exact-constraint})
are easily found, and the general case
can be understood as an interpolation between
these two limits [Fig.~\ref{fig:special_point_topology}].
The first limit is the XXZ0 limit~\cite{changlani18} $J_x=J_y=2/3$, $J_z=-1/3$
in which case $\mathcal{M}$ is composed of two spheres:
\begin{align}
\theta_1 &= \theta_2=\theta_3\,, \nonumber \\
\phi_1 &= \phi_2\pm\frac{2\pi}{3}=\phi_3\mp\frac{2\pi}{3}\,.
\label{eq:xxz0-wf}
\end{align}
These two spheres meet at the points $\theta_1 = 0$ and $\pi$ corresponding to the all-up and all-down states, as illustrated under Fig.~\ref{fig:special_point_topology}~a).

The other easily solved limit is the Ising limit: $J_x=J_y=0$, $J_z=1$.
In this case, product ground states have $\theta=0$ (spin fully up) on
one site and $\theta=\pi$ (spin fully down) on another, 
with $\theta, \phi$ on the remaining site completely free.
The product ground state manifold $\mathcal{M}$ is thus
composed of six spheres, each connected to
two other spheres at the points where
two spins have $\theta=0$ or $\theta=\pi$.
This is illustrated in Fig.~\ref{fig:special_point_topology}~b).
The six spheres correspond to the six ways of assigning an up spin, a down spin, and a free spin to the three sites of a triangle.
The surface of each individual sphere corresponds to the direction of the expectation value of the free spin.

The manifold for more general exactly solvable $\{ J \}$ interpolates between these two limits by smoothing out the singular points where the spheres meet so as to produce a manifold with the topology of a torus.
This interpolation can be made precise by
parameterizing the $\{ J \}$ satisfying~(\ref{eq:exact-constraint})
as $J_{x/y} = 1/3 - (2/3) \cos(\kappa \pm 2 \pi / 3)$,
$J_{z} = 1/3 - (2/3) \cos(\kappa)$. Then $\kappa=0$ is
a XXZ0 point [Fig.~\ref{fig:special_point_topology}~a)],
$\kappa = \pi/3$ an Ising point [Fig.~\ref{fig:special_point_topology}~b)],
and the $\mathcal{M}$ for generic $\kappa \in \langle 0, \pi/3 \rangle$
smoothly deforms as we vary $\kappa$ from one limiting case
to the other [Fig.~\ref{fig:special_point_topology}~c)].
In between, $\mathcal{M}$ has the topology of a torus,
and this torus pinches at two (six) points as $\kappa \to 0$ ($\kappa \to \pi/3$).
%

We have checked numerically that the
surface of single triangle solutions 
for general parameters has genus $g=1$
(see Appendix~\ref{app:topology}
and Supplemental Material~\cite{supplemental}).
This is itself an unusual situation,
which can be contrasted (e.g.) with the exact ground states of a Heisenberg ferromagnet
which cover the surface of a sphere.

Having obtained the exact solutions on a single triangle,
the question is then how much freedom there is
to tile these solutions over the whole lattice.

\subsection{Exact ground states on the lattice}
\label{subsec:gs_lattice}

Any $S=1/2$ product state [Eq.~(\ref{eq:product-wf})] can be labelled by
the expectation values of the spin operators
$\langle {\bf S}_i \rangle =
(\langle S^x_i \rangle, 
\langle S^y_i \rangle, \langle S^z_i \rangle)$,
up to a global phase.
Let us fix $\langle {\bf S}_i \rangle$
on one particular site and then seek to build
a solution on the lattice from there.

It turns out we can choose any direction for
the first $\langle {\bf S}_i \rangle$ and
still find configurations for the surrounding spins
such that the triangle and system are in
a ground state.
Specifying the first $\langle {\bf S}_i \rangle$ will
in general remove the  continuous freedom identified for the single triangle in~\ref{subsec:product_gs},
leaving only a discrete set of
possibilities for the neighboring spins to remain in
the ground state.

If we fix  $\langle {\bf S}_i \rangle$ to ${\bf S}_0$
and consider one of the two triangles connected to the
site $i$,
then the remaining spins $j,k$ on that triangle can
take only one of two configurations. These two configurations
are related to one another by
a permutation symmetry that swaps the remaining
pair of sites,
\begin{equation}
\langle {\bf S}_j \rangle = {\bf S}_1\,, 
\langle {\bf S}_k \rangle = {\bf S}_2
\ \ \text{or} \ \ 
\langle {\bf S}_j \rangle = {\bf S}_2\,, 
\langle {\bf S}_k \rangle = {\bf S}_1\,,
\end{equation}
with the values ${\bf S}_1$ and ${\bf S}_2$ being fixed
by ${\bf S}_0$.

\begin{figure}[t]
    \centering
    \includegraphics[width=0.3\columnwidth]{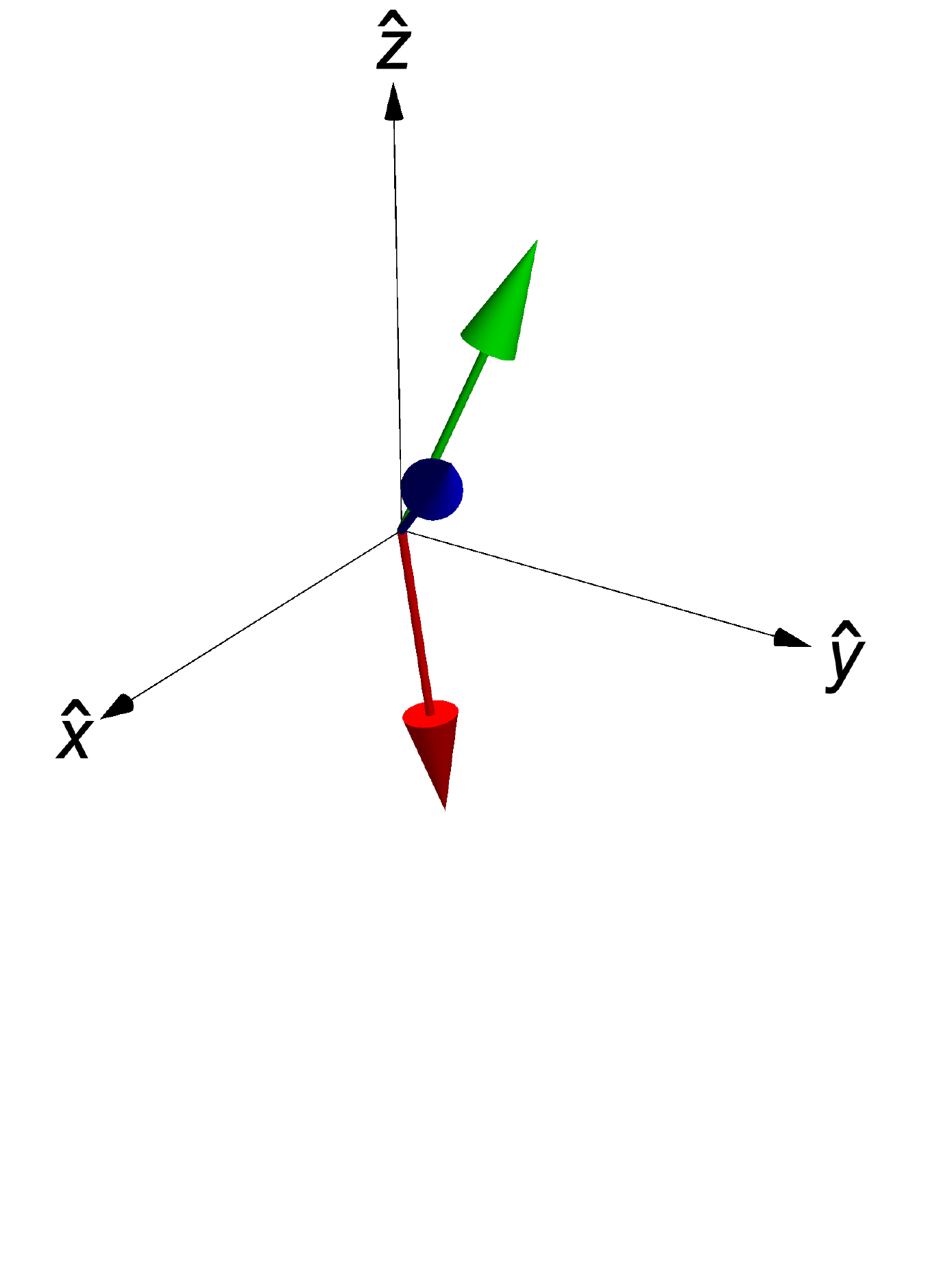}
    \ \ 
    \includegraphics[width=0.3\columnwidth]{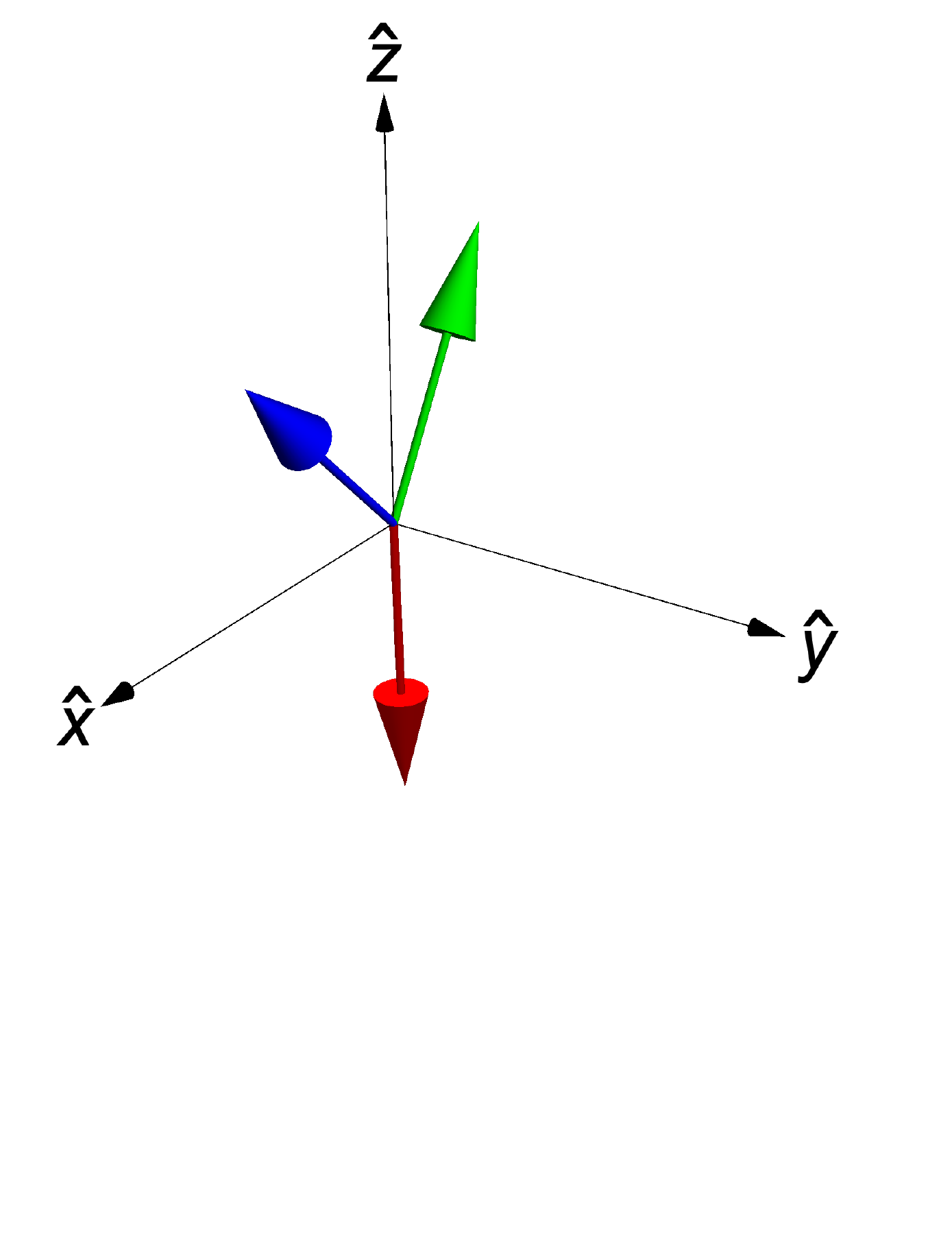}
    \ \ 
    \includegraphics[width=0.3\columnwidth]{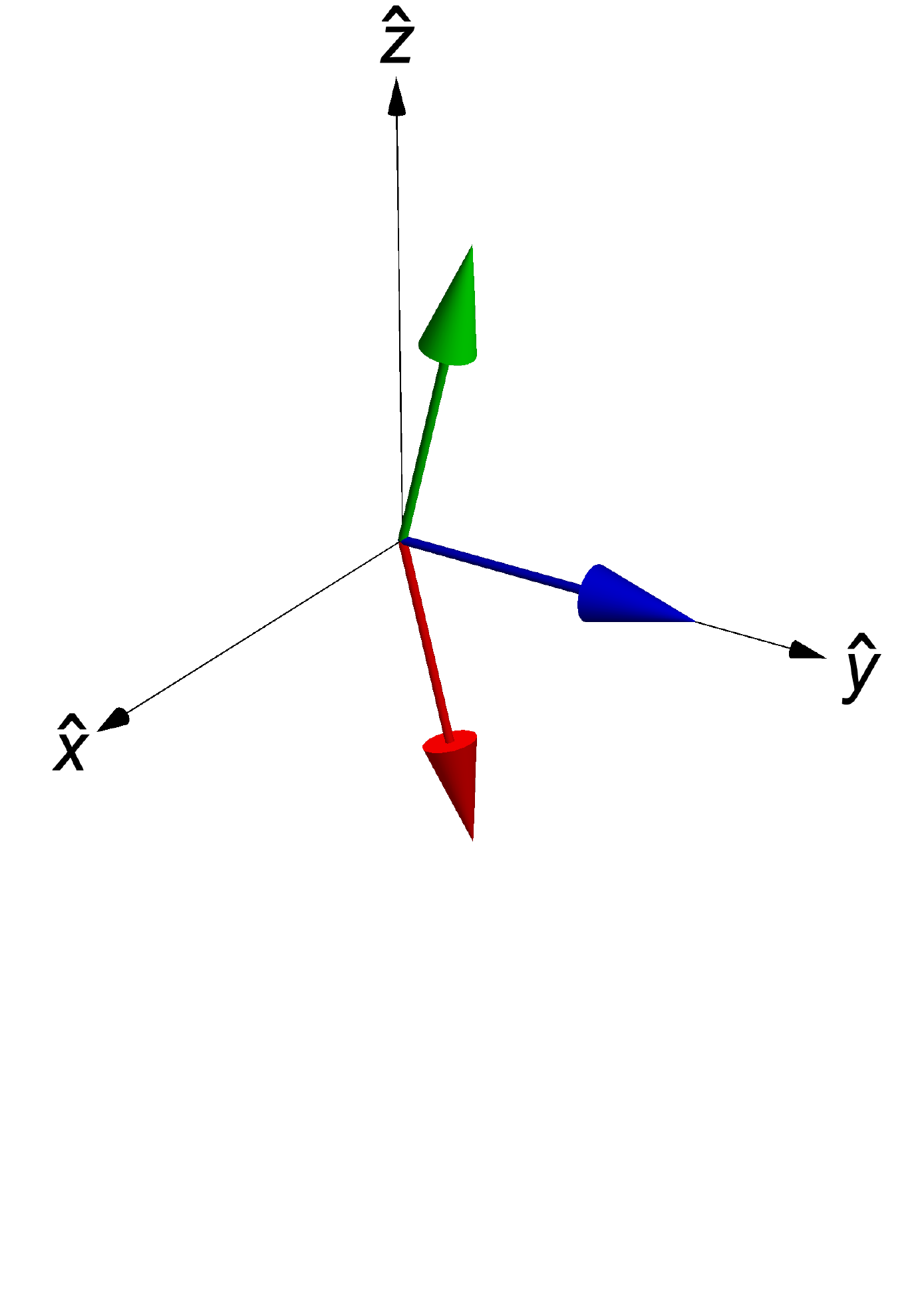}
    \caption{
    Examples of allowed 
    triads of spin expectation values which may occur
    together on a triangle, 
    for 
    an example parameter set
    $J_x/J_z=1/4$, $J_y/J_z=-1/5$.
    Different possible product ground state wavefunctions correspond to a different choice of one such triad on the first triangle,
    and then a tiling of the lattice with the three members of the triad,
    with each member occurring once on each triangle.
    Animations of all allowed spin
    triads for various parameter sets $\{J\}$ are included in the Supplementary Material~\cite{supplemental}.
    }
    \label{fig:gsproperties}
\end{figure}

Propagating this throughout the system,
consistency between triangles will force all
sites to take one of the three expectation
values $({\bf S}_0,{\bf S}_1,{\bf S}_2)$.
Some examples of allowed triads of expectation
values $({\bf S}_0,{\bf S}_1,{\bf S}_2)$ for
$J_x/J_z=1/4$, $J_y/J_z=-1/5$ are shown in 
Fig.~\ref{fig:gsproperties}.
Provided that the first triangle $ijk$ was in a
ground state, the permutation symmetry of 
$\mathcal{H}_{\text{XYZ}}$ guarantees that any triangle
where each of the three expectation values is represented once is also a ground state.

Provided that $({\bf S}_0,{\bf S}_1,{\bf S}_2)$ are all distinct from one
another (which is true for generic members of the ground state manifold),
these conditions are in precise correspondence to the
rules of the three coloring model.
Thus, the set
of product state solutions to $\mathcal{H}_{\text{XYZ}, \Lambda=0}$ is given by the space of three-color configurations,
along with two continuous degrees of freedom which were 
used up by fixing the first spin.

This establishes our main result: that $\mathcal{H}_{\text{XYZ}}$
with parameters chosen to 
obey Eq.~(\ref{eq:exact-constraint}) has a set of
exact, product, ground states, the number of which
grows exponentially with system size.
The wavefunctions described by Eq.~(\ref{eq:product-wf}) are not all linearly independent, so the actual ground state degeneracy is not the same as the number of product state solutions.
Nevertheless, if the number of product state solutions grows exponentially with system size, it should be expected that the number which are linearly independent also grows exponentially.
This was verified for the case of the XXZ0 model 
in Ref.~\onlinecite{changlani18}.

\begin{figure}[t]
    \centering
    \includegraphics[width=\columnwidth]{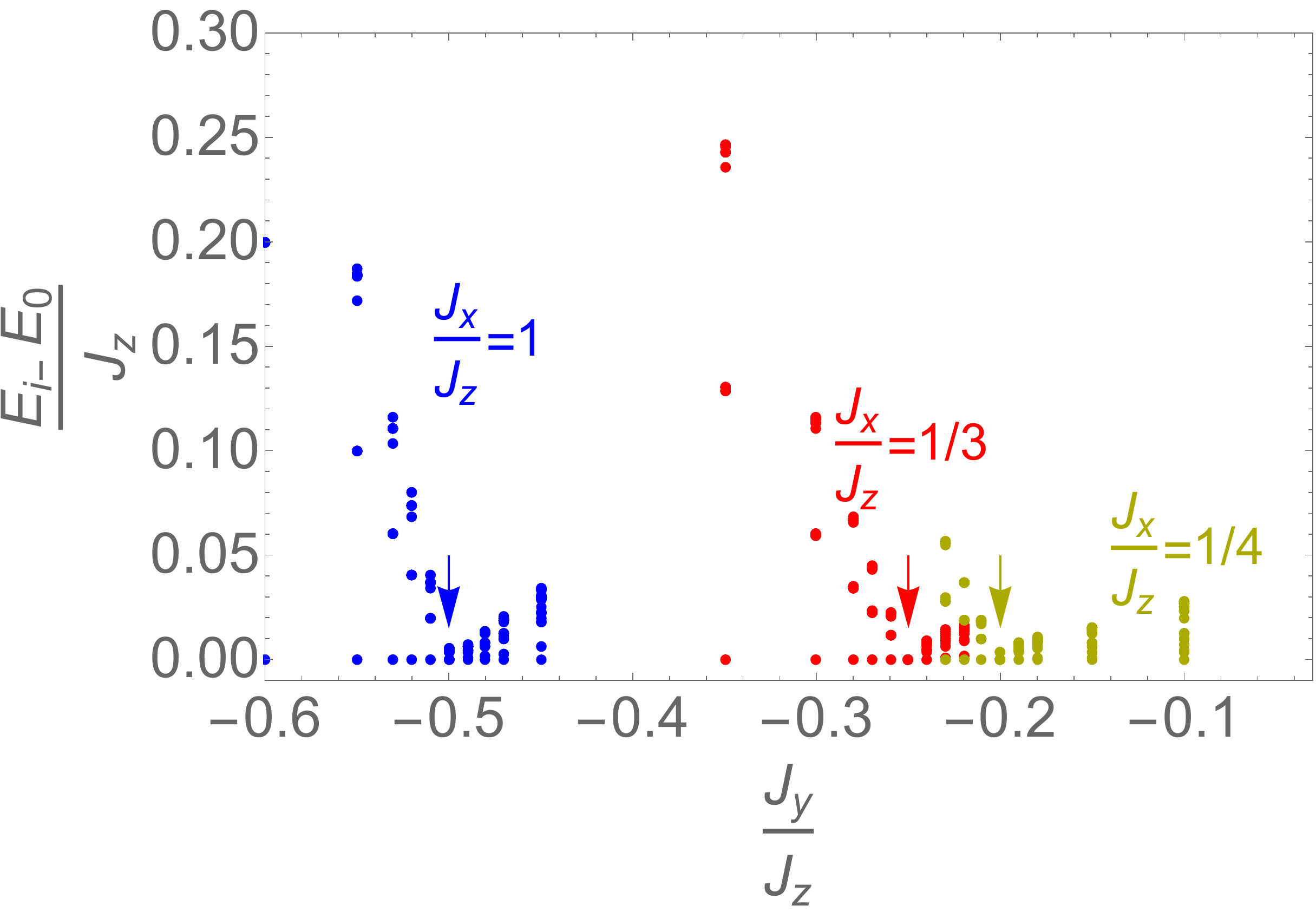}
    \caption{Low energy spectra of the XYZ model
    Eq.~(\ref{eq:XYZmodel}) 
    from exact diagonalization of a 24-site cluster, as a function
    of $J_y/J_z$ for $J_x/J_z=1$, $1/3$, $1/4$.
    A collapse of the excited states towards zero energy can be seen
    approaching the points where Eq.~(\ref{eq:exact-constraint}) is
    satisfied (respectively at $J_y/J_z=-1/2$, $-1/4$, $-1/5$ for the
    three chosen values of $J_x/J_z$, as indicated by arrows).
    }
    \label{fig:EDspectra}
\end{figure}

We have checked using exact diagonalization
on small clusters that there is indeed a collapse of many excited states towards the ground state approaching the points in parameter space given by (\ref{eq:exact-constraint}), consistent 
with the establishment of a macroscopic ground state degeneracy in the thermodynamic limit.
This is shown for a 24-site cluster in Fig.~\ref{fig:EDspectra}.

It is possible that there are also additional ground states, beyond the product ground states found here.
Changlani \textit{et al}.\ found numerically 
that there are indeed some additional ground states in
the XXZ0 limit for lattices with periodic
boundaries, although not with open boundaries~\cite{changlani18}.

The set of exactly solvable XYZ models described
by~(\ref{eq:exact-constraint}) includes both
the XXZ0 model ($\{J\}\propto(\frac{2}{3}, \frac{2}{3}, \frac{-1}{3})$ and permutations) and the Ising model ($\{J\}\propto(0,0,1)$ and permutations), 
three times each.
The Ising model represents a singular limit for the
set of product ground states because if two spins on a triangle are fixed to be one up and one down,
the third spin is actually completely undetermined,
increasing the freedom in the construction of  ground states on the lattice.

\section{Partial Ordering in the Ground States}
\label{sec:order}

In this Section we argue that the exact ground states
identified in Section~\ref{sec:model} possess infinite
range correlations, coexisting with the algebraic
correlations implied by the three-color mapping and are
in this sense partially ordered.

\begin{figure}
\centering
\includegraphics[width=0.7\columnwidth]{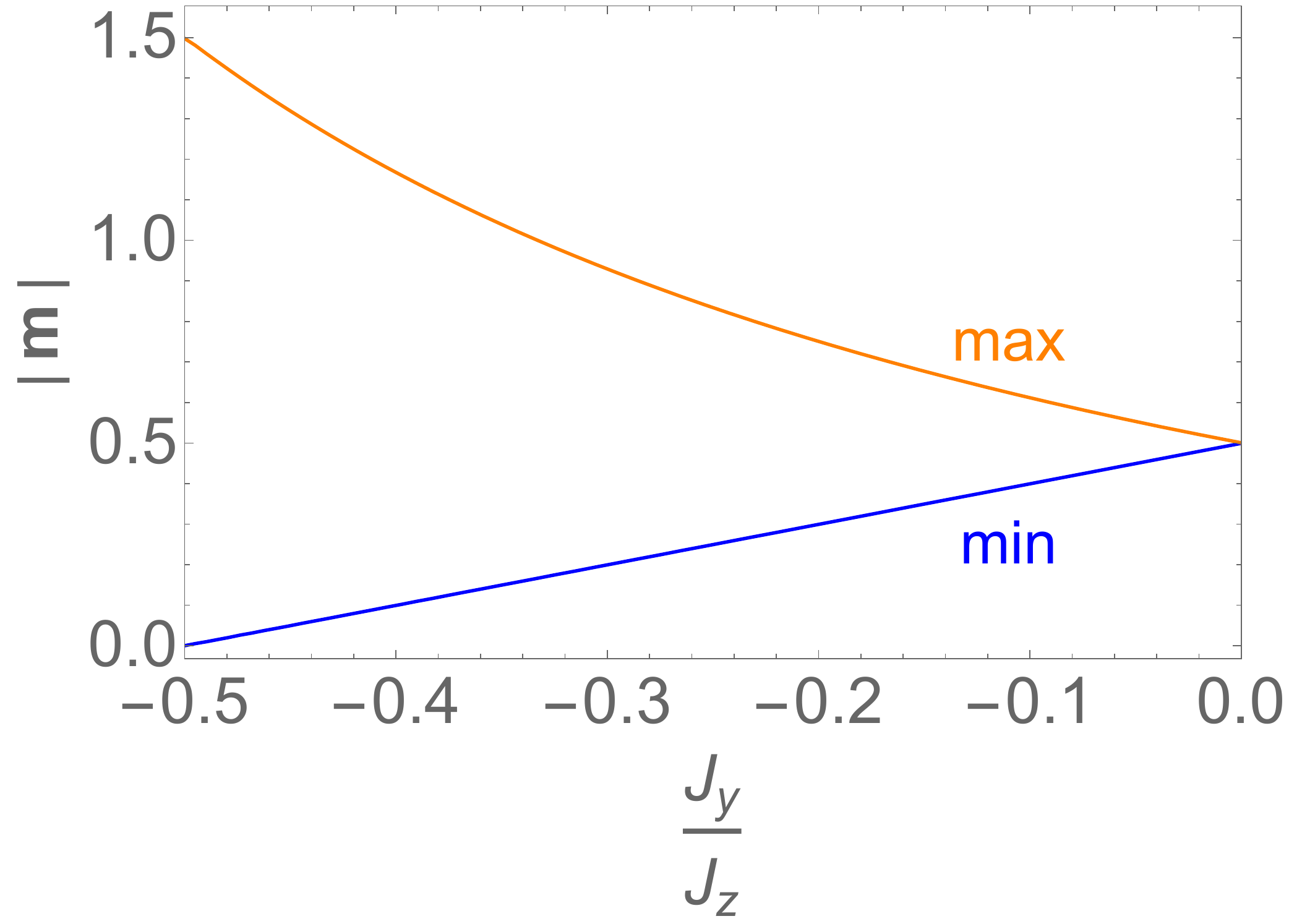}
\caption{
Maximum and minimum possible ground state values of $|{\bf m}|=|\langle {\bf S}_0 \rangle+\langle {\bf S}_1 \rangle+\langle {\bf S}_2 \rangle|$ within the set of exact
product ground states on a single triangle 
as a function of $J_y/J_z$,
for sets of parameters obeying condition (\ref{eq:exact-constraint}).
Apart from at the XXZ0 point ($J_y/J_z=-1/2$),
all allowed product ground states have a finite ${\bf m}$.
Once ${\bf m}$ is fixed on the first triangle it will be the same on all triangles throughout the lattice,
apart from in the Ising limit, ($J_y/J_z=0$) where some
individual spins can be rotated freely.
This establishes that, for general parameter sets obeying 
(\ref{eq:exact-constraint}), the ground states carry infinite range spin correlations, coexisting with the algebraic correlations implied by the three-color mapping.
}
\label{fig:finitemag}
\end{figure}

The allowed ground state configurations of
spin expectation values on a single triangle
generally have a finite value of 
${\bf m}=\langle {\bf S}_0 \rangle+\langle {\bf S}_1 \rangle+\langle {\bf S}_2 \rangle$.
This is illustrated in 
Fig.~\ref{fig:finitemag}, where the minimum and maximum
value of $|{\bf m}|$ over the possible single
triangle ground states is plotted as a function 
of $J_y/J_z$.
Note that ${\bf m}$ would not be the same as
the magnetisation in a real system, because of the
local basis used to define the spins [Fig.~\ref{fig:kagome_3colors}].

Since the remaining triangles in the lattice must only feature
permutations of the spin expectation values on the first triangle (see Section~\ref{subsec:gs_lattice}), and since ${\bf m}$ is invariant under those
permutations, we conclude that all triangles must have the same value of ${\bf m}$. 
This implies that ${\bf M}=\sum_{\triangle} {\bf m}_{\triangle}$  has a macroscopic value in any given product state and that
there are infinite range correlations of the
spin component parallel to ${\bf m}$.
These infinite-range correlations must 
coexist with the algebraic 
correlations that are also present due to the mapping between
the spin configurations and the three-color model.

\section{Gapless Excitations}
\label{sec:excitations}

In addition to the many ground states, this exactly
solvable model also possesses gapless excitations.

We demonstrate this below by making use of the continuous degeneracy of the ground states identified in Section~\ref{subsec:product_gs}, and a trial wave function for the excitations based on the single mode approximation.
Starting from 
a translationally invariant member of the set of ground states
one can construct excitations
using the generators of rotations within the
single triangle ground state manifold, and show that these are gapless in the long wavelength limit.

Let us emphasize that the considerations
of this section depend only on
the continuous degeneracy of the ground states, and not
on the discrete degeneracy of how one can tile the one triangle
solution across the whole lattice. Thus the argument of this
section would work equally well for (e.g.) exactly solvable XYZ models on the triangular lattice, which would retain the continuous degeneracy of the kagome case considered here, but for which only 6 
consistent
three-colorings are possible.

The gapless excitations are similar to Nambu-Goldstone modes,
except that unlike ordinary Nambu-Goldstone
modes the associated symmetry generator $\mathcal{G}$
does not commute with the Hamiltonian
\begin{equation}
[\mathcal{H}_{\text{XYZ}}, \mathcal{G}] \neq 0\,,
\end{equation}
but rather the commutator annihilates a ground state:
\begin{equation}
[\mathcal{H}_{\text{XYZ}}, \mathcal{G}] \lvert \Psi_0 \rangle =0\,. \label{eq:quasi-sym-gen-commut}
\end{equation}
One way of motivating the above is to consider a
one-parameter family of ground states $\lvert  \Psi_t \rangle$
that we know exists because of the continuous degeneracy.
Then it is always possible to find an unitary operator $\mathcal{U}_t$
that maps $\lvert \Psi_0 \rangle \mapsto \lvert \Psi_t \rangle$.
We may then identify the quasisymmetry generator 
with $\mathcal{G} = i \partial_t \mathcal{U}_t|_{t=0}$.
Since $E_0 \lvert \Psi_1 \rangle = \mathcal{H}_{\text{XYZ}} \mathcal{U}_t \lvert \Psi_0 \rangle = \mathcal{U}_t \mathcal{H}_{\text{XYZ}} \lvert \Psi_0 \rangle$, (\ref{eq:quasi-sym-gen-commut}) follows.
Qualitatively, we may say that $\mathcal{G}$ becomes an exact symmetry only in the zero-energy limit.

Of course, it may be the case that some of the
ground state quasi-symmetry generators are also
genuine symmetry generators. The XXZ0 point, considered
in Section~\ref{subsec:product_gs}, is a good example.
From~(\ref{eq:xxz0-wf}) it is evident that a global rotation
around the $\hat{\bf z}$ axis maps a ground state to a
ground state. In fact, this is an exact symmetry of the XXZ Hamiltonian.
On the other hand,
rotating the spins by $\theta_i \mapsto \theta_i + \delta \theta$
also maps a ground state to a ground state, but is not
a genuine symmetry.
Note how in the case of the $\theta_i \mapsto \theta_i + \delta \theta$
transformation, the generator $\mathcal{G}$ depends on the starting
ground state $\lvert \Psi_0 \rangle$.

To begin with, we choose a member of the set of product ground states to construct excitations around.
In such a ground state,
each spin expectation value takes one of
three values $\langle {\bf S}_r \rangle$, $\langle {\bf S}_g \rangle$, and $\langle {\bf S}_b \rangle$,
according to whether
the site is red, green, or blue 
in the three-color representation
of the state. 
$\lvert \Psi_0 \rangle$ is chosen to be the state corresponding to
a translationally invariant
three-coloring of the lattice.
Later we comment on more
general red-green-blue patterns.

The existence of a continuous two-dimensional manifold of exact product ground states on the single triangle implies the existence of infinitesimal global rotations
which keep every triangle, and therefore the system as a whole, in a ground state.

There are two independent
generators of these rotations for every starting ground state (spin triad). We will pick one
of them, which we label $\mathcal{G}=\sum_{i} g_{i}$. The site generator $g_{i}$ depends on the color of site $i$ in the three-color tiling of the ground state.

We now do a site-dependent 
coordinate transformation on the spin basis $\{ x,y,z \} \to 
\{ u_i,v_i,w_i \}$, with the new coordinate axes chosen such that ${\bf \hat w}_i$ aligns with the expectation value of spin $i$ in the ground state
and so that the generators $g_{i}$ are
proportional to rotations
around ${\bf \hat u}_i$.
The spin operators in this basis then satisfy:
\begin{align}
S^w_i \lvert \Psi_0 \rangle &= \frac{1}{2}\lvert  \Psi_0 \rangle\,, \quad \forall \, i\,,
\\
g_i &= U_i \, S^u_i\,, \label{eq:U-scaling-factors}
\end{align}
where $U_i$ is a real scaling factor
that depends only on the coloring
of the site $i$.
(In general, $U_i$ also depend on the
three spin vectors $\langle {\bf S}_r \rangle$, $\langle {\bf S}_g \rangle$, and $\langle {\bf S}_b \rangle$,
but these are the same across $\lvert \Psi_0 \rangle$.)
It is normalized according to
\begin{equation}
\sum_{c = r,g,b} (U_c)^2 = 1\,.
\end{equation}

Since the local ${\bf \hat u}_i$ axes have been chosen 
to correspond with the rotation
axes that keep the system in a
ground state, it must be true that:
\begin{equation}
\sum_i [\mathcal{H}_{\text{XYZ}}, U_i S^u_i] \lvert  \Psi_0 \rangle = 0\,.
\label{eq:comm_annihilate}
\end{equation}

We then consider the following variational
wavefunction for the excitations,
based on the single mode approximation,
\begin{align}
\lvert {\rm ex}, {\bf q}\rangle &=
 \mathcal{G}({\bf q})\lvert \Psi_0 \rangle\,,
\label{eq:exq-def} \\
\mathcal{G}({\bf q}) &=
 \frac{1}{\sqrt{N}}
\sum_{i} U_i S^u_i \exp(i {\bf q} \cdot {\bf r}_i)\,,
\label{eq:su-def}
\end{align}
where $N$ is the number of unit cells.
The wavefunction $\lvert {\rm ex}, {\bf q}\rangle$ describes a spin wave-like excitation.

$\lvert {\rm ex}, {\bf q}\rangle$ is orthogonal to $\lvert \Psi_0 \rangle$
because $\langle \Psi_0 \vert  S^u_i \vert \Psi_0 \rangle$ vanishes everywhere.
At finite ${\bf q}$ and in the thermodynamic limit $N\to\infty$,
it is also orthogonal to the other members of the ground state manifold.
Because of this, the expectation value of the energy in $\lvert {\rm ex}, {\bf q}\rangle$ is an upper bound on the energy of excitations with momentum ${\bf q}$:
\begin{equation}
E(\mathbf{q})-E_0 \leq \frac{\langle \Psi_0 \vert
\mathcal{G}(-{\bf q}) [\mathcal{H}_{\text{XYZ}}, \mathcal{G}({\bf q})]
\vert \Psi_0 \rangle
}{\langle \Psi_0 \vert \mathcal{G}(-{\bf q}) \mathcal{G}({\bf q}) \vert \Psi_0 \rangle}\,.
\label{eq:SMA1}
\end{equation}

The denominator $\langle \Psi_0 \vert \mathcal{G}(-{\bf q}) \mathcal{G}({\bf q}) \vert \Psi_0 \rangle=\frac{1}{4}$, and we can set $E_0=0$ (i.e., measure all energies relative to the ground state), so that
\begin{equation}
E(\mathbf{q}) \leq 4 \langle \Psi_0 \vert
\mathcal{G}(-{\bf q}) [\mathcal{H}_{\text{XYZ}}, \mathcal{G}({\bf q})] 
\vert \Psi_0 \rangle\,.
\label{eq:SMA2}
\end{equation}

Due to Eq.~(\ref{eq:comm_annihilate}),
the ${\bf q}\to {\bf 0}$ limit of Eq.~(\ref{eq:SMA2}) vanishes:
\begin{equation}
\lim_{{\bf q}\to {\bf 0}} E(\mathbf{q})=0\,,
\end{equation}
implying gapless excitations.

The part of Eq.~(\ref{eq:SMA2}) which is linear in ${\bf q}$
also vanishes:
\begin{align}
&\frac{1}{N} \sum_{m,n} i{\bf q} \cdot {\bf r}_m
\langle \Psi_0 \vert
[U_n S^u_n, \mathcal{H}_{\text{XYZ}}] U_m S^u_m 
\vert \Psi_0 \rangle \nonumber \\
&- \frac{1}{N} \sum_{m,n} i{\bf q} \cdot {\bf r}_n
\langle \Psi_0 \vert
U_n S^u_n [\mathcal{H}_{\text{XYZ}}, U_m S^u_m] 
\vert \Psi_0 \rangle = 0\,,
\end{align}
which also follows from Eq.~(\ref{eq:comm_annihilate}).

This leads us to the conclusion that the dispersion of
excitations 
has a quadratic upper-bound
at small $q$:
\begin{equation}
E({\bf q}) \leq \zeta \, q^2\,.
\end{equation}

This also agrees with a linear spin wave analysis
which finds
quadratically dispersing excitations
around the translationally
invariant exact ground states.

Although not obvious, a detailed
analysis of the properties
of the coupling matrix $J_{ij}$
in the new basis $\{u_i,v_i,w_i\}$
shows that the second generator of symmetry
has $g_i' = U_i S^v_i$ with the same
scaling factors $U_i$ from
Eq.~(\ref{eq:U-scaling-factors}).
Thus one finds that the commutator
$[\mathcal{G}, \mathcal{G}']$ has
a non-vanishing expectation value
in the ground state $\vert \Psi_0 \rangle$,
implying that the two generators
represent only one degree of freedom
(cf.\ $[x, p] = i \hbar$)~\cite{Watanabe-NGcounting}.
In the spin wave analysis this is
reflected in the fact that there is
only one gapless mode,
despite the two broken quasi-symmetry
generators.

For more general coloring patterns
(that are not translationally symmetric),
$\mathcal{G}({\bf q})$ may still be
used to probe the low-lying excitations,
but this time ${\bf q}$ cannot be identified
with the momentum.
The above argument
thus suggests that
low-lying excitations still exist
even for more general RGB patterns.

With that said, one should
keep in mind that the
above argument may
fail if the upper bound from (\ref{eq:SMA1})
is discontinuous at ${\bf q} = {\bf 0}$,
(see e.g.\ Supplementary Material of~[\onlinecite{BCS-argument-counterexample}]).

To verify that our upper bound is continuous,
we evaluate it by noting that only
averages of the form
$\langle \Psi_0 \vert S_i^{u} S_i^{\mu} S_{i+\delta}^{w} \vert \Psi_0 \rangle$
for $\mu = u,v$ are non-vanishing, giving:
\begin{eqnarray}
&&\hspace{-6pt}\langle \Psi_0 \vert
\mathcal{G}(-{\bf q}) [\mathcal{H}_{\text{XYZ}}, \mathcal{G}({\bf q})] 
\vert \Psi_0 \rangle
= \nonumber \\
&&\hspace{-6pt}- \frac{1}{8 N} \sum_{i \delta} (U_i)^2 J^{i,i+\delta}_{ww}
+ \frac{1}{8 N} \sum_{i \delta} U_i U_{i + \delta} e^{i {\bf q} \cdot {\bf \delta}} (J^{i,i+\delta}_{vv} - i J^{i,i+\delta}_{uv})\,. \nonumber \\[-8pt]
\label{eq:inter-step-dbl}
\end{eqnarray}

From the fact that $\vert \Psi_0 \rangle$
is an exact eigenstate, it follows that in
the new basis the exchange coefficients
satisfy $J^{ij}_{uu} = J^{ij}_{vv} = J^{ij}_{\perp}$
and $J^{ij}_{uv} = - J^{ij}_{vu} = d^{ij}$.
The exchange coefficients $J^{ij}_{\mu\nu}$ also depend only on
the coloring of the sites $i$ and $j$ and satisfy
$J^{ij}_{\mu\nu} = J^{ji}_{\nu\mu}$.
Moreover, from the fact that the change in energy
within one triangle is to second
order equal to zero when we vary the
spins with $\mathcal{G}$,
it follows that:
\begin{eqnarray}
&&2 J_{\perp}^{rg} U_r U_g
+ 2 J_{\perp}^{gb} U_g U_b
+ 2 J_{\perp}^{br} U_b U_r  - (J_{ww}^{rg} + J_{ww}^{rb}) (U_r)^2 \nonumber \\
&&
- (J_{ww}^{gr} + J_{ww}^{gb}) (U_g)^2
- (J_{ww}^{br} + J_{ww}^{bg}) (U_b)^2 = 0\,.
\end{eqnarray}

Applying to Eq.~(\ref{eq:inter-step-dbl}), we obtain:
\begin{eqnarray}
&&\hspace{-6pt}\frac{\langle \Psi_0 \vert
\mathcal{G}(-{\bf q}) [\mathcal{H}_{\text{XYZ}}, \mathcal{G}({\bf q}) ]
\vert \Psi_0 \rangle
}{\langle \Psi_0 \vert \mathcal{G}(-{\bf q}) \mathcal{G}({\bf q}) \vert \Psi_0 \rangle} = \nonumber \\
&&\hspace{-6pt}\frac{1}{4 N} \sum_{\langle i j\rangle}
U_i U_j \left[
J^{ij}_{\perp} [\cos 
\left( {\bf q} \cdot ({\bf r}_j  - {\bf r}_i)
\right)- 1]
+ d^{ij} 
\sin \left( {\bf q} \cdot ({\bf r}_j  - {\bf r}_i)\right)
\right], \nonumber \\[-8pt]
\end{eqnarray}
which is indeed continuous and vanishing in the
limit ${\bf q} \to {\bf 0}$.

We therefore conclude that the partially ordered exact ground states of the
XYZ model have gapless excitations, despite the absence of continuous symmetry
in the original Hamiltonian.

\section{Recipe for constructing further exactly solvable models}
\label{sec:recipe}

Our work exposes a simple recipe for the construction of
further highly degenerate exactly solvable models, of the same kind as those discussed here.

First, one must define a Hamiltonian for quantum spins of length $S$ on a set of corner sharing units (for instance, triangles or tetrahedra), with the Hamiltonian on each unit being symmetric under permutations of the sites.
Then one must tune the parameters such that the  degeneracy $d_0$ of the ground state of the single unit Hamiltonian is large enough that product wavefunctions on the single unit have enough free parameters to be made orthogonal to all of the excited states.

If $n$ is the number of sites in each unit then, the total number of states in the single unit spectrum is \mbox{$(2S+1)^n$} 
and a single unit product state has $4Sn$ degrees of freedom.
Requiring that the real and imaginary parts of the overlap with every excited state in the single unit spectrum vanishes gives $2((2S+1)^n-d_0)$ constraints, so to be able to find product-like solutions we need:
\begin{eqnarray}
d_0\geq(2S+1)^n-2Sn\,.
\end{eqnarray}
If the single unit Hamiltonian can be tuned to have a sufficiently large $d_0$ then one can search for product-like ground states.
If, in these ground states, all $n$ sites in the unit are distinguishable from one another (e.g., if the spin expectation values differ on the sites), and the single-unit Hamiltonian has a permutation symmetry under the swapping of sites, then the problem of tiling solutions over the whole lattice becomes an $n$-coloring problem.
We anticipate that this recipe can be used to construct further exactly solvable models on other lattices.

In the case of $S=1/2$ spins on the kagome 
lattice, the minimal $d_0$ is $5$ (cf.\ $d_0=6$ for the models considered in this manuscript).
In a model with $d_0=5$,
there would be no continuous degrees of freedom left
in the ground state,
but only discrete degrees of freedom
from the three-coloring pattern.

The exactly solvable ``two-coloring'' models identified in 
Ref.~\onlinecite{pal-arxiv} can also be seen as an example of the construction described above, with the corner-sharing units
being single bonds ($n=2$) and the requirement for exact solvability $d_0\geq2$. The ``two-coloring'' models in Ref.~\onlinecite{pal-arxiv} have $d_0=3$.

\section{Summary and outlook}
\label{sec:summary}

We have demonstrated that the exactly solvable model
pointed out in~Ref.~\onlinecite{changlani18} is a member of a wider
family of exactly solvable models, and that the ground states
of these models possess coexisting infinite-range
and algebraic correlations.
The combination of the large ground state degeneracy,
and the coexistence of 
infinite-range
and algebraic correlations
suggest an analogy with the 
phenomenon of
``magnetic moment fragmentation''~\cite{brooks14,rougemaille19,lhotel20,paddison16,canals16,petit16,benton16,lefrancois17}
but the case here is distinguished by the fact that 
it occurs in the exact ground states of a quantum model.

Our conclusions generalize straightforwardly to XYZ models on
other triangle-based lattices, such as the hyperkagome lattice.

The models described here, fall into a wider category of
frustrated systems possessing a large number of low energy
states, defined by local constraints. 
In such cases, a description of the low energy physics in 
terms of emergent gauge fields and charges is frequently
useful~\cite{isakov04, moessner10, cepas11}, and this may be an interesting avenue to investigate
further for the present case.

The exact ground states we have defined are not themselves
quantum spin liquids, since they lack quantum entanglement and have a large non-topological degeneracy.
Nevertheless, highly degenerate points of a model are often a good starting point for discovery of spin liquids because small perturbations can stabilize a variety of non-trivial superpositions of the degenerate states.
In this case, given the partial ordering of the ground states,
perturbing the models identified here may be a way to stabilize
phases which combine the interesting features of a quantum spin liquid (entanglement, fractional excitations) with spontaneous symmetry breaking.
With this in mind, we suggest that a numerical study
of the ground states of $\mathcal{H}_{\text{XYZ}}$ for parameter
sets close to the exactly solvable limit may be a
rewarding subject for future work.

In view of the large number of ground states, and low lying excited states, the physics of the exactly solvable models
at finite temperature is also an interesting 
topic for the future.
In particular, it is possible that at $T>0$
thermal order-by-disorder will lead to a breaking 
of the ground state degeneracy. 
Quantum ($T=0$) order-by-disorder is conclusively ruled out
for the models described in this work, because the exact ground states and their energies are known.
It may nevertheless be the case that thermal fluctuations can distinguish between the states, leading to an entropic
selection at finite temperature.
The low energy gapless modes discussed in Section~\ref{sec:excitations} are likely to play a key role in this selection, if it occurs.

A further unusual feature of the XYZ 
model presented here is that even the solutions to the single triangle problem are topologically non-trivial, in that the continuous manifold of single-triangle solutions has the topology of a torus.
Whether interesting phenomena can be derived from this by, for example, adiabatically 
moving the system around this manifold is something which remains to be seen.

Finally we note that these exact ground states may also be of interest from the point of view of non-equilibrium physics.
With a simple alteration of the Hamiltonian~\cite{lee20}
the exact ground states can be turned into many-body quantum scars: highly excited states which violate the eigenstate thermalization hypothesis.
The XYZ models proposed here offer a chance to explore this in a setting which lacks continuous spin rotation symmetry.

\appendix

\section{Relationship between XYZ model and more general anisotropic exchange models on the kagome lattice}
\label{app:gen_models}

\begin{figure}[b]
    \centering
    \includegraphics[width=0.9\columnwidth]{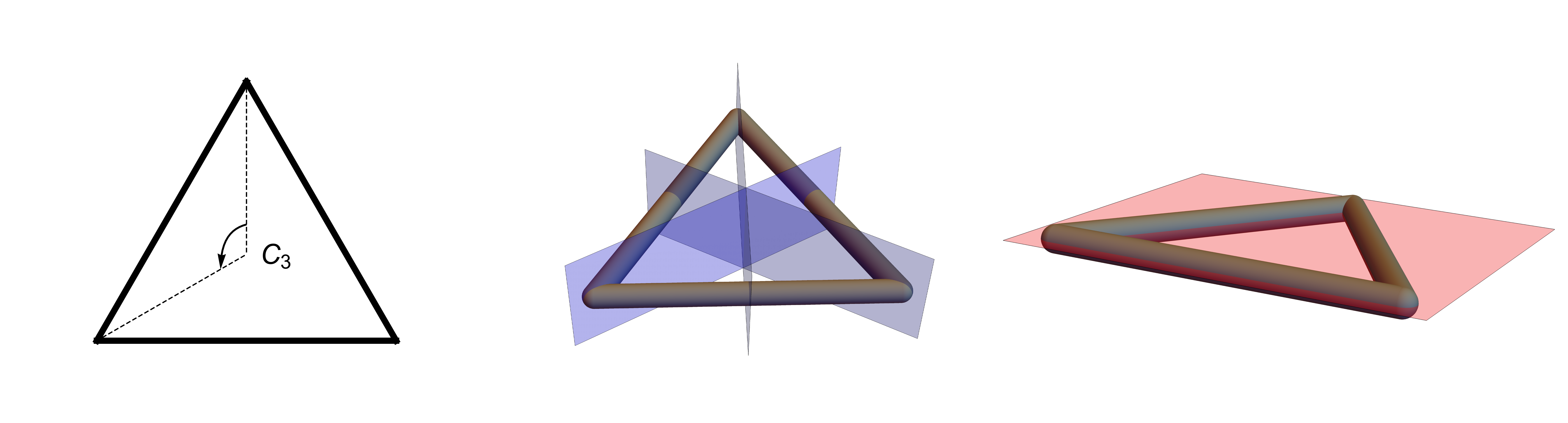}
    \caption{Symmetries of a triangle in the kagome lattice used to constrain the exchange Hamiltonian, Eq.~(\ref{eq:H_glob})~\cite{Essafi2017}.
    These are: $C_3$ rotations around the center of the triangle, reflections in the planes perpendicular to each bond, and reflection in the plane of the lattice itself.
    }
    \label{fig:tri_symm}
\end{figure}

Here we elaborate on the relationship
between the XYZ model studied in
the main text and the model dicussed in Ref.~\onlinecite{Essafi2017}.
The model discussed in Ref.~\onlinecite{Essafi2017} 
gives the most general set of nearest neighbour
interactions consistent with the symmetries 
of the lattice, including a mirror symmetry 
in the kagome plane itself.
In total, the symmetries considered comprise the
inversion and translation symmetries of the kagome lattice along with the three point group symmetries shown in Fig.~\ref{fig:tri_symm}.

\begin{figure}[b]
\centering
\includegraphics[width=0.6\columnwidth]{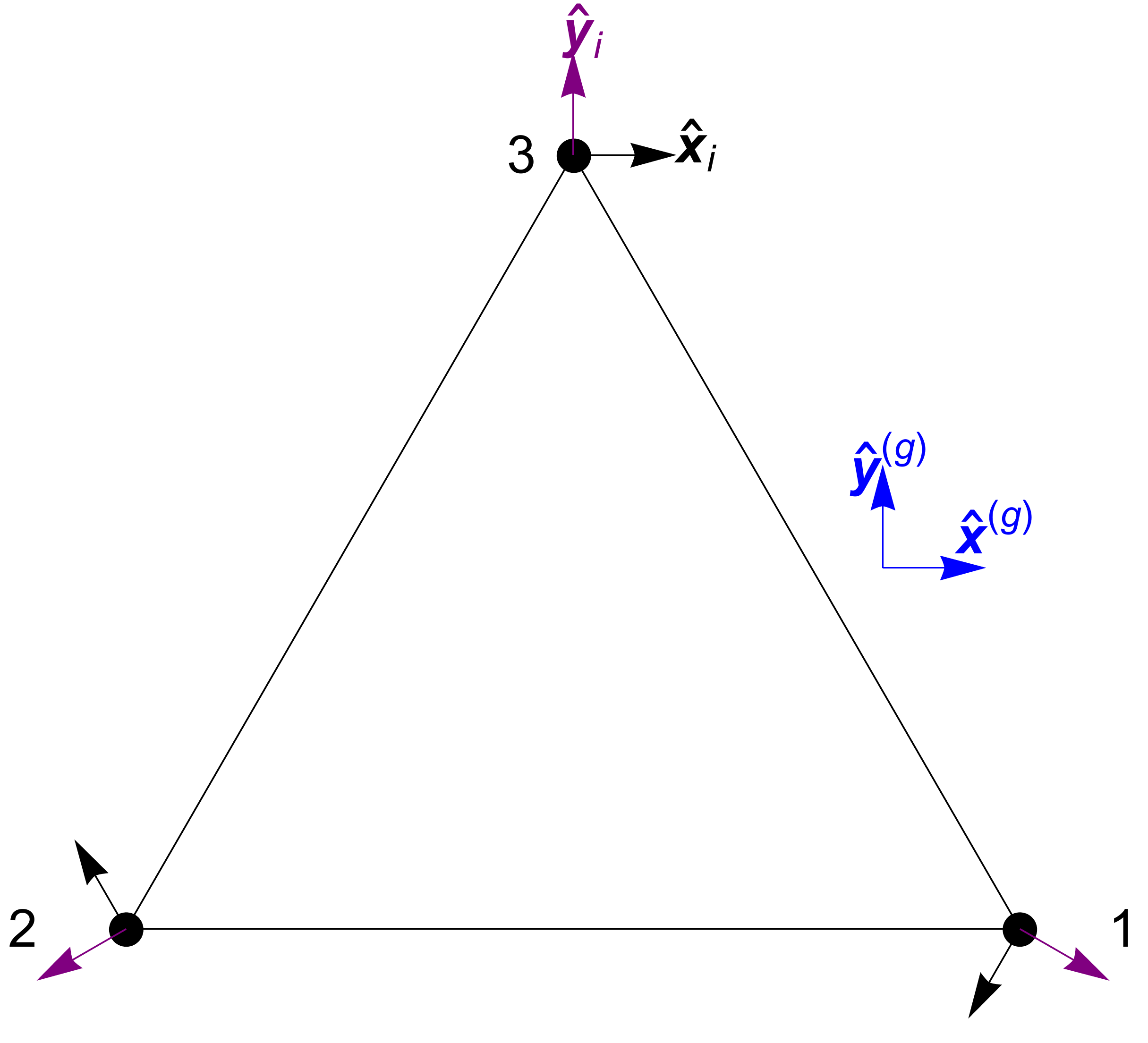}
\caption{Global and local bases used in Eqs.~(\ref{eq:H_glob})-(\ref{eq:H_loc}), and numbering
convention used for sites on a triangle.}
\label{fig:localbasis}
\end{figure}

The Hamiltonian of this model is:
\begin{eqnarray}
\mathcal{H}_{\sf gen}=
\sum_{\langle ij \rangle}
S^{\alpha}_i J^{\alpha \beta}_{ij} S^{\beta}_j\,.
\end{eqnarray}
The three coupling matrices on the triangle, when written in
the global basis $( \hat{\bf{x}}^{(g)},\hat{\bf{y}}^{(g)}, \hat{\bf{z}})$ [Fig.~\ref{fig:localbasis}], have the form:
\begin{eqnarray}
\hspace{-1cm}
&&J_{12}=\begin{pmatrix}
J_{\perp}+t & D_z & 0 \\
-D_z & J_{\perp}-t & 0 \\
0 & 0 & J_z
\end{pmatrix}\,, \nonumber \\
&&J_{23}=\begin{pmatrix}
J_{\perp}-\frac{1}{2}t & D_z+\frac{\sqrt{3}}{2}t & 0 \\
-D_z+\frac{\sqrt{3}}{2}t & J_{\perp}+\frac{1}{2}t & 0 \\
0 & 0 & J_z
\end{pmatrix}\,, \nonumber \\
&&J_{31}=\begin{pmatrix}
J_{\perp}-\frac{1}{2}t & D_z-\frac{\sqrt{3}}{2}t & 0 \\
-D_z-\frac{\sqrt{3}}{2}t & J_{\perp}+\frac{1}{2}t & 0 \\
0 & 0 & J_z
\end{pmatrix}\,,
\label{eq:H_glob}
\end{eqnarray}
with the numbering convention for sites on a triangle as shown
in Fig.~\ref{fig:localbasis}.

Transforming to a local coordinate system, with local
axes in the $xy$-plane ${\bf \hat x}_i,{\bf \hat y}_i $
on each site of the triangle as shown in Fig.~\ref{fig:localbasis}, while maintaining the same global 
${\bf \hat z}$ axis, the coupling matrices become uniform:
\begin{eqnarray}
&&\hspace{-4pt} \tilde{J}= 
R_i \cdot J_{ij} \cdot R^T_j=
\nonumber \\
&&\hspace{-4pt} \begin{pmatrix}
-\frac{1}{2}\left( 
\sqrt{3} D_z + J_{\perp}
\right) +t
& \frac{1}{2}\left( 
 -D_z + \sqrt{3} J_{\perp}
\right) & 0 \\
 \frac{1}{2}\left( 
 D_z - \sqrt{3} J_{\perp}
\right) & -\frac{1}{2}\left( 
\sqrt{3} D_z + J_{\perp}
\right) -t  & 0 \\
0 & 0 & J_z
\end{pmatrix}. \nonumber \\[-8pt]
\label{eq:H_loc}
\end{eqnarray}

In the special case $D_z=\sqrt{3} J_{\perp}$,
this becomes an XYZ model with
\begin{eqnarray}
J_x=-2 J_{\perp} + t\,, \quad J_y=-2 J_{\perp} -t\,.
\end{eqnarray}

This establishes that the XYZ model discussed in the main text occurs as a limit of the general symmetry-allowed anisotropic exchange Hamiltonian for
kagome magnets.

\section{Derivation of the exact ground states}
\label{app:derivation}

Here we show how to obtain analytic expressions for
the ground states in the exactly solvable limits
of the XYZ model.

The ground state wave functions are product states
of the form Eq.~(\ref{eq:product-wf})
Such a wave function is completely determined (up to a global phase) by the
expectation values of the spin components on each site:
\begin{eqnarray}
&&\langle S^x_i \rangle=\frac{1}{2} \sin(\theta_i) \cos(\phi_i)\,, \nonumber \\ 
&& \langle S^y_i \rangle=\frac{1}{2} \sin(\theta_i) \sin(\phi_i)\,,  \nonumber \\ 
&&\langle S^z_i \rangle=\frac{1}{2}  \cos(\theta_i).
\label{eq:exp_val}
\end{eqnarray}
The vector $\langle {\bf S}_i \rangle=(
\langle S^x_i \rangle, 
\langle S^y_i \rangle,
\langle S^z_i \rangle)$ is constrained to lie on a
sphere:
\begin{eqnarray}
\langle S^x_i \rangle^2+
\langle S^y_i \rangle^2+
\langle S^z_i \rangle^2=1/4\,, \quad \forall \, i\,.
\label{eq:lengthconstraint}
\end{eqnarray}

In the following, we find analytic expressions for $\langle {\bf S}_i \rangle$ for all product state ground states of the single triangle Hamiltonian, at the exactly solvable points of the XYZ model. The ground states on the lattice follow from this by tiling the lattice with single triangle solutions in the way described in the main text.

The exactly solvable parameter
space is given by $J_x J_y + J_y J_z + J_z J_x=0$ and
$J_x+J_y+J_z>0$. 
Setting $J_x+J_y+J_z=1$ as the unit of energy
we can then write the exchange parameters
as a function of a single variable $\kappa$:
\begin{align}
J_{x/y} &= \frac{1}{3}+\frac{1}{3}\cos(\kappa)
\pm\frac{1}{\sqrt{3}}\sin(\kappa)\,, \nonumber \\
J_z &= \frac{1}{3}-\frac{2}{3}\cos(\kappa)\,. \label{eq:kappadef}
\end{align}
The points $\kappa = 0$, $2\pi/3$, and $4\pi/3$ correspond to the XXZ0 points along the ${\bf \hat z}$, ${\bf \hat y}$, and ${\bf \hat x}$ axis, respectively,
whereas $\kappa = \pi/3$, $\pi$, and $5\pi/3$ correspond to the Ising points along the ${\bf \hat x}$, ${\bf \hat z}$, and ${\bf \hat y}$ axis, respectively.
If we attribute $\beta = x \to 1$, $y \to 2$, and $z \to 3$, then the above can be
compactly written as:
\begin{eqnarray}
J_{\beta} = \frac{1}{3} - \frac{2}{3} \cos\left(\kappa + \frac{2 \pi}{3} \beta\right)\,. \label{eq:compactkappadef}
\end{eqnarray}

The above parameterisation has the nice property that it allows
us to focus on the parameter range $\kappa \in [0, \pi/3]$,
with the rest related to this range by a permutation
of the $\{J\}$ coefficients and spin components. In particular:
\begin{eqnarray}
(J_x, J_y, J_z)\big|_{- \kappa} = (J_y, J_x, J_z)\big|_{\kappa}\,,
\label{eq:Jcoef-perm1} \\
(J_x, J_y, J_z)\big|_{\kappa+2\pi/3} = (J_y, J_z, J_x)\big|_{\kappa}\,.
\label{eq:Jcoef-perm2}
\end{eqnarray}
This permutation symmetry (of the $x,y,z$ components)
relates systems with different exchange coefficients and should
not be conflated with the dynamical permutation symmetry of
exchanging the spins on a triangle, whilst keeping $\{J\}$ the same.
The latter we used in Section~\ref{subsec:gs_lattice}
to derive the tiling rules.

Since the XXZ0 and Ising points were analyzed
in the main text
under Section~\ref{subsec:product_gs}, below
we focus on the case of generic $\kappa \in \langle 0, \pi/3 \rangle$.

\subsection{Method of Lagrange multipliers}

For generic values of
$\kappa$, not equal to $\mathbb{Z} \pi / 3$,
we find the ground state by minimising the expectation value of the energy, using Lagrange
multipliers to enforce Eq.~(\ref{eq:lengthconstraint}).
If this expectation value $\langle \Psi \vert \mathcal{H}_{\text{XYZ}} \vert \Psi \rangle$
coincides with the known ground state energy
$e_0=-\frac{1}{4}(J_x+J_y+J_z)=-1/4$,
then $\lvert \Psi \rangle$ is a ground state.

The energy of the product state 
can be written directly in terms of
the spin expectation values:
\begin{equation}
E_{\Psi}= \langle \Psi \vert \mathcal{H}_{\text{XYZ}} \vert \Psi \rangle =
\sum_{\langle ij \rangle}
\sum_{\beta=x,y,z} J_{\beta} \langle S^{\beta}_i \rangle\langle S^{\beta}_j \rangle\,.
\end{equation}
To simplify the notation, we will from
this point forward denote $\langle S^{\beta}_i \rangle$ as $S^{\beta}_i$ and treat them as classical variables.
The problem is then to minimize $E_{\Psi}$ with respect to the
nine variables $S^{\beta}_i$ (three components $\beta$ on
each of three sites of a triangle $i$), while
respecting the three constraints Eq.~(\ref{eq:lengthconstraint}).
This can be achieved
with the method of Lagrange multipliers.

We seek to minimize:
\begin{align}
&\begin{aligned}
E' = \frac{1}{4} &+ \frac{1}{2} \sum_{i \neq j} \sum_{\beta=x,y,z}
J_{\beta} S_{i}^{\beta} S_{j}^{\beta} \\
&- \frac{1}{2} \sum_i \mu_{i} \left( {\bf S}_i^2-\frac{1}{4} \right) = 0\,,
\end{aligned} \\
&\frac{\partial E'}{\partial S_i^{\beta}}=0\,, \quad \forall \, i,\beta\,,
\label{eq:min1} \\
&\frac{\partial E'}{\partial \mu_i}=0, \quad \forall \, i\,,
\label{min2}
\end{align}
where we have introduced three Lagrange multipliers $\mu_i$ 
to the energy, to satisfy  Eq.~(\ref{eq:lengthconstraint}).

Eq.~(\ref{eq:min1}) has the form of zero-eigenvalue
equation for a block diagonal matrix
\begin{align}
{\mathcal M}\cdot\tilde{S} &= 0\,, 
\label{eq:MS}
\\
\tilde{S} &= \left(
S^x_1 \ 
S^x_2 \ 
S^x_3 \ 
S^y_1 \ 
S^y_2 \ 
S^y_3 \ 
S^z_1 \ 
S^z_2 \ 
S^z_3
\right)^{T}, \label{eq:Stilde}
\\
\mathcal{M} &=
\begin{pmatrix}
M_x & 0 & 0 \\
0 & M_y & 0 \\
0 & 0 & M_z
\end{pmatrix},
\  
M_{\beta}=\begin{pmatrix}
- \mu_1 & J_{\beta} & J_{\beta} \\
J_{\beta} & - \mu_2 & J_{\beta} \\
J_{\beta} & J_{\beta} & - \mu_3
\end{pmatrix}. \nonumber \\
\end{align}

For there to be solutions 
of Eq.~(\ref{eq:MS}) where all spin components are finite on at least one spin in the
triangle, we require that:
\begin{eqnarray}
\det M_{\beta}= 2 J_{\beta}^3 +
 J_{\beta}^2( \mu_1+\mu_2+\mu_3)
 - \mu_1 \mu_2 \mu_3 = 0\,,
 \quad \forall \, \beta\,. \nonumber \\
 \label{eq:detM}
\end{eqnarray}

Acting on Eq.~(\ref{eq:MS}) from the left with
$\tilde{S}$, and using the spin length constraints
and the known ground state energy, gives us
\begin{eqnarray}
\mu_1+\mu_2+\mu_3=-2(J_x+J_y+J_z)=-2\,.
\label{eq:sum-mu}
\end{eqnarray}

Substituting Eq.~(\ref{eq:sum-mu}) in to Eq.~(\ref{eq:detM}),
gives the relation:
\begin{eqnarray}
\mu_1 \mu_2 \mu_3 =
2 J_{\beta}^2 (J_{\beta} - 1)
=
- \frac{8}{27} \cos^2\left( \frac{3 \kappa}{2} \right)\,. 
\label{eq:prod-mu}
\end{eqnarray} 
Eqs.~(\ref{eq:sum-mu}) and (\ref{eq:prod-mu}) are sufficient
to guarantee the satisfaction of all three conditions in Eq.~(\ref{eq:detM}).

The next step is then to solve Eqs.~(\ref{eq:sum-mu}) and (\ref{eq:prod-mu}) to give expressions for $\mu$ in terms of $\kappa$.
To do this, we write $\mu_j$ in terms of polar coordinates, an angle $u$ and radius $r(u)$,
\begin{eqnarray}
\mu_j=-\frac{2}{3}-\frac{2}{3}r(u) \cos\left( u + \frac{2\pi j}{3} \right)\,.
\label{eq:mu-polar}
\end{eqnarray}
This solves Eq.~(\ref{eq:sum-mu}), and  Eq.~(\ref{eq:prod-mu}) now
gives a cubic equation for $r(u)$:
\begin{eqnarray}
\cos(3u) \, r^3 - 3 \, r^2
+ 4 \sin^2\left( \frac{3\kappa}{2} \right) = 0\,. \label{eq:cubic-r}
\end{eqnarray}

For general $u \neq \pi/6 + \mathbb{Z} \pi/3$, that is $\cos(3u) \neq 0$,
the above equation has three real roots
$r_n(u,\kappa)$, $n=0,1,2$, given by  
Vi\`{e}te's formula:
\begin{eqnarray}
&& r_n(u) = \sec(3u) - 2|\sec(3u)|\cos\left(\Phi(u)+
\frac{2\pi n}{3}
\right)\,, 
\\
&&\hspace{-10pt} \Phi(u)=\frac{1}{3} {\rm arccos}\left[
{\rm sgn}(\cos(3u))
\left[
-1+2\cos^2(3u)\sin^2\left(3\kappa/2\right)
\right]
\right]\,. \nonumber \\
\end{eqnarray}
The identities ${\rm arccos}(1 - 2 x^2) = 2 |{\rm arcsin}(x)|$ and ${\rm arccos}(x) + {\rm arccos}(-x) = \pi$, when combined with a piecewise redefinition of $n$, allow us to further simplify the above to:
\begin{eqnarray}
&&\hspace{-20pt} r_n(u) = \sec(3u)\left[1 + 2 \cos\left[\frac{2}{3} {\rm arcsin}(\cos(3u) \sin(3\kappa/2)) + \frac{2\pi n}{3}\right]\right]\,.
\nonumber \\
 \label{eq:r-solution}
\end{eqnarray}
Since we're using polar coordinates to express $\mu_j$ in Eq.~(\ref{eq:mu-polar}), $(r,u)$ and $(-r,u+\pi)$ represent
the same $\mu_j$ and the $n=1$ and $n=2$ radial solutions are equivalent. Thus there are only two distinct solutions,
having $n=0$ and $1$.

Some properties of $r_{n}(u,\kappa)$ and $\mu_j(u,n,\kappa)$:
\begin{itemize}
\item[(i)] Since $r_n(u,\kappa)$
depends on $u$ only through $\cos(3u)$, it
satisfies $r_n(u,\kappa) = r_n(-u,\kappa) = r_n(u + 2\pi/3,\kappa)$
and therefore $\mu_j$ satisfies
$\mu_j(u,n,\kappa) = \mu_{j+3}(u,n,\kappa) = \mu_{j+1}(u-2\pi/3,n,\kappa) = \mu_{-j}(-u,n,\kappa)$.
Hence results derived for $\mu_3$ transfer to $\mu_1$ and $\mu_2$
(i.e., $\mu_1$ and $\mu_2$ do not need to be considered separately).

\item[(ii)] Only for the XXZ0 points,
$\kappa = \mathbb{Z} 2 \pi/3 \implies \sin(3\kappa/2) = 0$,
does Eq.~(\ref{eq:cubic-r})
have a (double) solution $r_{1/2} = 0$.
For all other $\kappa \neq \mathbb{Z} 2 \pi/3$
and $u \in \mathbb{R}$, $r_{n}(u,\kappa) \neq 0$.
The other solution at the XXZ0 points is $r_0 = 3 / \cos(3u)$.

\item[(iii)] At the Ising points,
$\kappa = \pi/3 + \mathbb{Z} 2 \pi/3 \implies \sin^2(3\kappa/2) = 1$,
the three solutions of Eq.~(\ref{eq:cubic-r})
can be written as $r^{{\rm Ising}}_n(u) = - 1 / \cos(u + 2 \pi n /3)$.
A piecewise patching of the different $n$ solutions from Eq.~(\ref{eq:r-solution})
is needed to construct these smooth $r^{{\rm Ising}}_n$.

\item[(iv)] By studying the limit $u \to \pi/6+\mathbb{Z}\pi/3$
of $r_{n}(u)$ where $\sec(3u)$ diverges,
one may confirm that the $n=0$ solution of Eq.~(\ref{eq:r-solution})
diverges there, whereas $r_{n=1} = - (2 / \sqrt{3}) \sin(3 \kappa / 2)$
at these points. The latter value also follows directly from Eq.~(\ref{eq:cubic-r}).
As for $\mu_j$, the $\cos(u+2\pi j/3)$ from Eq.~(\ref{eq:mu-polar}) eliminates
some of the $n=0$ divergences. Specifically, for $\mu_3(n=0)$ the only
unremovable poles are at $u = \pm \pi/6$ and $u = \pm 5 \pi/6$ (for all $\kappa$).

\item[(v)] At the Ising points $\kappa = \pi/3 + \mathbb{Z} 2 \pi/3$,
${\rm arcsin}(\cos(3u))$ is a
sawtooth-like function that is not smooth at $u = \mathbb{Z} \pi/3$.
Consequently, $r_{n=0}(u)$ as given by Eq.~(\ref{eq:r-solution}) is not
smooth at $(u, \kappa) = (\mathbb{Z}\pi/3,\pi/3+\mathbb{Z}2\pi/3)$,
and $r_{n=1}(u)$ is not smooth
at $(u, \kappa) = (\pi/3+\mathbb{Z}2\pi/3,\pi/3+\mathbb{Z}4\pi/3)$
and $(\mathbb{Z}2\pi/3,\pi+\mathbb{Z}4\pi/3)$.
For all other $\kappa$,
the functions
$r_n(u,\kappa)$ and $\mu_j(u,n,\kappa)$ are smooth in $u$
on their
whole domain of definition (which for $n=0$ excludes
certain singularities).
Likewise, for fixed $u = \mathbb{Z} \pi/3$, ${\rm arcsin}(\sin(3\kappa/2))$ from
the $r_n(\kappa)$ are not smooth at the same $(u,\kappa)$ points from above.
For all other fixed $u$,
the functions
$r_n(u,\kappa)$ and $\mu_j(u,n,\kappa)$ are smooth in $\kappa$
for all $\kappa \in \mathbb{R}$.

\item[(vi)] Lastly, for a fixed $n$, $r_n(u,\kappa)$ and $\mu_j(u,n,\kappa)$
as functions of $(u,\kappa)$ are smooth everywhere on their domains
of definitions, excluding the cusps listed under (v). This is a stronger statement
than being smooth in only $u$ or $\kappa$, and it follows from the
structure of Eqs.~(\ref{eq:r-solution}) and (\ref{eq:mu-polar}).
With that said,
in the arguments that follow we shall
only be needing the piecewise continuity and differentiability in $u$ of these
functions.
\end{itemize}


With the values of $\mu_j$ determined, the vector of
spin components $\tilde{S}$ [Eq.~(\ref{eq:Stilde})]
is given by a linear combination of the null vectors
of $\mathcal{M}$.
Row reduction of
$M_{\beta}$ gives
\begin{eqnarray}
&&\hspace{-18pt}M_{\beta} \sim \begin{pmatrix}
J_{\beta} + \mu_1 & - (J_{\beta} + \mu_2) & 0 \\
0 & J_{\beta} + \mu_2 & - (J_{\beta} + \mu_3) \\
- \mu_1 & J_{\beta} & J_{\beta} \\
\end{pmatrix} \nonumber \\
&&\sim \begin{pmatrix}
J_{\beta} + \mu_1 & - (J_{\beta} + \mu_2) & 0 \\
0 & J_{\beta} + \mu_2 & - (J_{\beta} + \mu_3) \\
0 & 0 & 0 \\
\end{pmatrix},
\label{eq:M-beta-row-reduction}
\end{eqnarray}
where in the second step we have used Eq.~(\ref{eq:detM})
and assumed that at least two $i$ out of three satisfy
$J_{\beta} + \mu_i \neq 0$.
It turns out that for all $\kappa$, there are special values of
$u$, denoted $u^*$,
for which two $J_{\beta} + \mu_i$ (with the same $\beta$) vanish,
invalidating the second step above.

For generic $\kappa \neq \mathbb{Z}\pi/3$
(i.e., not Ising or XXZ0 points), however, these special $u^*$
are restricted to a discrete number of points.
Although not obvious, all $\tilde{S}$ solutions of these special $u^*$ can
be obtained by a limiting procedure of the generic $u$ solutions.

Below we first determine when $J_{\beta} + \mu_i = 0$ and find
that these special $u^*$ are, for generic $\kappa$, given by $u^* = \mathbb{Z}\pi/3$.
Then we find the spins $\tilde{S}$
at these special $u^*$ explicitly
to confirm that there are no
points isolated from the generic $u$ solutions.
Lastly,
we find the $\tilde{S}$ for generic $u$.

\subsection{Finding the special $u^*$}

In the coming two sections, the functions $J_{\beta} + \mu_i$
shall play a prominent role, so we introduce
\begin{eqnarray}
a_{\beta i}(u, n, \kappa) = \frac{1}{2}
\big(J_{\beta}(\kappa) + \mu_i(u, n, \kappa)\big)\,.
\label{eq:a-semi-axis-def}
\end{eqnarray}

Special $u^*$ we define as those $u$ for which
least one $a_{\beta i}(u^*)$ vanishes. Let us consider
the case where $a_{x1} = 0$. Then $\mu_1 = - J_x$
and one may easily solve Eqs.~(\ref{eq:sum-mu}) and (\ref{eq:prod-mu})
to get $(\mu_2 + J_x) (\mu_2 + 2 - 2 J_x) = 0$ and $\mu_3 = J_x - 2 - \mu_2$.
Thus $\mu_1 = - J_x$ implies that either
$\mu_2 = -J_x$ and $\mu_3 = 2 (J_x - 1)$,
or $\mu_2 = 2 (J_x - 1)$ and $\mu_3 = - J_x$.
Focusing on the case $\mu_1=\mu_2=-J_x$ and $\mu_3 = 2 (J_x - 1)$,
we see that $a_{x1} = a_{x2} = 0$ and $a_{x3} = 3 (J_x - 2/3) \neq 0$
when we're not at a XXZ0 point.
We may therefore
conclude that $a_{\beta i}$ always vanish in pairs (with the same $\beta$),
but for generic $\kappa$ never in triplets (i.e.,
$a_{\beta 1} = a_{\beta 2} = a_{\beta 3} = 0$
never happens).

These special points $\mu_1=\mu_2=-J_x$ and $\mu_3 = 2 (J_x - 1)$, moreover,
coincide with the extrema of $\mu_3$.
To prove this, we differentiate
Eqs.~(\ref{eq:sum-mu}) and (\ref{eq:prod-mu}) to get, for all $u$,
\begin{align}
\partial_u \mu_1 + \partial_u \mu_2 + \partial_u \mu_3 &= 0\,,
\label{diff-r-eq1} \\
(\partial_u \mu_1) \mu_2 \mu_3
+ \mu_1 (\partial_u \mu_2) \mu_3 
+ \mu_1 \mu_2 (\partial_u \mu_3) &= 0\,. \label{diff-r-eq2}
\end{align}
By substituting $\mu_1=\mu_2=-J_x$, $\mu_3 = 2 (J_x - 1)$ in to the above,
one obtains $J_x (J_x - 2/3) \partial_u \mu_3(u^*) = 0$.
Since $J_{\beta} = 0$ only at the Ising points, and $J_{\beta} = 2/3$
only at the XXZ0 points, we conclude that
for generic $\kappa$, $\partial_u \mu_3(u^*) = 0$.

The converse statement also holds. That is, $\partial_u \mu_3(u^*) = 0$
implies that for at least one $\beta$,
$\mu_1(u^*)=\mu_2(u^*)= - J_{\beta}$ and
$\mu_3(u^*) = 2 (J_{\beta} - 1)$.
To prove this, we use $\partial_u \mu_3(u^*) = 0$ in
Eqs.~(\ref{diff-r-eq1}) and (\ref{diff-r-eq2}) to
obtain $(\mu_1 - \mu_2) \mu_3 \partial_u \mu_1 = 0$.
The case $\mu_3 = 0$ is forbidden for non-Ising $\kappa$
because of Eq.~(\ref{eq:prod-mu}). The case
$\partial_u \mu_1 = \partial_u \mu_2 = \partial_u \mu_3 = 0$
happens only for the XXZ0 point with $r=0$.
The generic case $\mu_1 = \mu_2$, when combined with
Eqs.~(\ref{eq:sum-mu}) and (\ref{eq:prod-mu}), yields
$(\mu_1 + J_x) (\mu_1 + J_y) (\mu_1 + J_z) = 0$
and $\mu_3 = - 2 \mu_1 - 2$, which
is the desired result.

Using $\partial_u \mu_3(u^*) = 0$ one can find the special $u^*$
without solving any complicated algebraic equations
that include the expression (\ref{eq:r-solution}).
First, we differentiate the cubic equation (\ref{eq:cubic-r})
and solve for $\partial_u r$,
\begin{eqnarray}
\frac{\partial_u r}{r} = \frac{r \sin(3u)}{r \cos(3u) - 2}\,.
\end{eqnarray}
Next, we differentiate the definition (\ref{eq:mu-polar}) of $\mu_j$
and use $\partial_u \mu_3(u^*) = 0$ to get
\begin{eqnarray}
\frac{\partial_u r(u^*)}{r(u^*)} = \frac{\sin(u^*)}{\cos(u^*)}\,.
\end{eqnarray}
By combining the above two equations, one obtains the desired
equation
\begin{eqnarray}
(r(u^*) \cos(u^*) + 1) \sin(u^*) = 0\,.
\end{eqnarray}
The only case when $r(u^*) = - 1 / \cos(u^*)$ is a solution
of the cubic equation (\ref{eq:cubic-r}) is
at the Ising points. Thus,
for generic $\kappa$, $a_{\beta 1} = a_{\beta 2} = 0$
occurs only at $u^* = \mathbb{Z} \pi$. Analogously,
$a_{\beta 1} = a_{\beta 3} = 0$ occurs
only at $u^* = 2\pi/3 + \mathbb{Z} \pi$,
and
$a_{\beta 2} = a_{\beta 3} = 0$ occurs
only at $u^* = \pi/3 + \mathbb{Z} \pi$.
Altogether, the special $u^*$ are given by $\mathbb{Z} \pi/3$.

These special $u^*$ also arise when finding
the $\tilde{S}$. The limiting condition that arises in that
context is $|a_{\beta i}| = |a_{\beta j}|$. Focusing on the
case $i=1, j=2$, the question is when is
$|a_{\beta 1}| = |a_{\beta 2}|$ for some $\beta$?
To answer this, one must consider two cases,
$a_{\beta 1} = + a_{\beta 2} \implies \mu_1 = \mu_2$
and $a_{\beta 1} = - a_{\beta 2} \implies 2 J_{\beta} + \mu_1 + \mu_2 = 0$.
In either case, after solving
with Eqs.~(\ref{eq:sum-mu}) and (\ref{eq:prod-mu}),
one finds that $\mu_1=\mu_2=-J_{\beta'}$ and $\mu_3 = 2 (J_{\beta'} - 1)$,
for some potentially different $\beta'$.
Thus $|a_{\beta 1}| = |a_{\beta 2}|$ happens only
at $u^* = \mathbb{Z} \pi$, and more generally
$|a_{\beta i}| = |a_{\beta j}|$
occurs only at $u^* = \mathbb{Z} \pi/3$.
Indeed,
when we know that, say, $|a_{x1}| = |a_{x2}|$,
then $|a_{y1}| = |a_{y2}|$ and $|a_{z1}| = |a_{z2}|$
must hold as well, and only one of these three vanishes,
the others being finite;
$|a_{x 3}|, |a_{y 3}|, |a_{z 3}|$ also must be non-zero.
All of this holds for generic $\kappa \neq \mathbb{Z}\pi/3$.

\subsection{Ground states for special $u^* = \mathbb{Z}\pi/3$ and $0 < \kappa < \pi/3$}

Let us consider a special point $u^*$
where $|a_{\gamma i}| = |a_{\gamma j}| = 0$
for a fixed $n=0,1$.
The remaining component and spin indices we shall denote
$\beta_1, \beta_2$ and $k$, respectively,
so that $(\beta_1,\beta_2,\gamma)$ and $(i,j,k)$
are permutations of $(x,y,z)$ and $(1,2,3)$,
respectively. 
We moreover consider only the range $0 < \kappa < \pi/3$,
since the parameterisation
permutation symmetry [Eqs.~(\ref{eq:Jcoef-perm1})-(\ref{eq:Jcoef-perm2})]
maps this region to all other generic $\kappa$.

From the previous section we know that
$|a_{\beta_1 i}| = |a_{\beta_1 j}| \neq 0$,
$|a_{\beta_2 i}| = |a_{\beta_2 j}| \neq 0$, and
$|a_{\beta_1 k}|, |a_{\beta_2 k}|, |a_{\gamma k}| \neq 0$
are all non-zero. In light of Eq.~(\ref{eq:M-beta-row-reduction}),
the null vectors of $M_{\beta_1}$ and $M_{\beta_2}$ are
given by
\begin{eqnarray}
V_{\beta}=2\begin{pmatrix}
(J_{\beta}+\mu_1)^{-1} \\
(J_{\beta}+\mu_2)^{-1} \\
(J_{\beta}+\mu_3)^{-1}
\end{pmatrix}=\begin{pmatrix}
a_{\beta 1}^{-1} \\
a_{\beta 2}^{-1} \\
a_{\beta 3}^{-1}
\end{pmatrix}\,,
\label{eq:regular-null-vec}
\end{eqnarray}
whereas the null vector of $M_{\gamma}$ has the
components $(V_{\gamma})_i = - (V_{\gamma})_j = 1$
and $(V_{\gamma})_k = 0$.

Since $\tilde{S}$ is a null vector of $\mathcal{M}$ [Eq.~(\ref{eq:MS})],
we may write it as:
\begin{eqnarray}
\tilde{S} = \frac{1}{2} \begin{pmatrix}
X_x V_x \\
X_y V_y \\
X_z V_z
\end{pmatrix}\,,
\label{eq:tilde-S-null-expr}
\end{eqnarray}
where the coefficients $X_{\beta}$ are
fixed by the spin normalization
constraints [Eq.~(\ref{eq:lengthconstraint})].

Because of the relations among $\{a_{\beta i}\}$, only
two of the three spin normalization
constraints are independent, to wit
\begin{eqnarray}
\frac{X^2_{\beta_1}}{|a_{\beta_1 i}|^2}
+ \frac{X^2_{\beta_2}}{|a_{\beta_2 i}|^2}
+ X^2_{\gamma} &=& 1\,, \label{eq:special-u-spin-norm1} \\
\frac{X^2_{\beta_1}}{|a_{\beta_1 k}|^2}
+ \frac{X^2_{\beta_2}}{|a_{\beta_2 k}|^2} &=& 1\,.
\label{eq:special-u-spin-norm2}
\end{eqnarray}
Eq.~(\ref{eq:special-u-spin-norm1}) defines an ellipsoid
and Eq.~(\ref{eq:special-u-spin-norm2}) a cylinder.
These two intersect at:
\begin{eqnarray}
&&X_{\beta_1}(v) = a_{\beta_1 k} \cos(v)\,, \label{eq:spec-param1} \\
&&X_{\beta_2}(v) = a_{\beta_2 k} \sin(v)\,, \label{eq:spec-param2} \\
&&X_{\gamma}(v) = \pm \sqrt{1
- \left(\frac{a_{\beta_1 k}}{a_{\beta_1 i}}\right)^2 \cos^2(v)
- \left(\frac{a_{\beta_2 k}}{a_{\beta_2 i}}\right)^2 \sin^2(v)
}\,, \nonumber \\ \label{eq:spec-param3}
\end{eqnarray}
and the spins are given by
\begin{gather}
{\bf S}_{i} =  \frac{1}{2} \begin{pmatrix}
X_{\beta_1} / a_{\beta_1 i} \\
X_{\beta_2} / a_{\beta_2 i} \\
+ X_{\gamma}
\end{pmatrix}\,, \ \ 
{\bf S}_{j} =  \frac{1}{2} \begin{pmatrix}
X_{\beta_1} / a_{\beta_1 j} \\
X_{\beta_2} / a_{\beta_2 j} \\
- X_{\gamma}
\end{pmatrix}\,, \nonumber \\
{\bf S}_{k} =  \frac{1}{2} \begin{pmatrix}
X_{\beta_1} / a_{\beta_1 k} \\
X_{\beta_2} / a_{\beta_2 k} \\
0
\end{pmatrix}\,,
\end{gather}
where the rows represent the $\beta_1$, $\beta_2$,
and $\gamma$ components of the spins.

What these equations represent depends on the
ratios $(a_{\beta_1 k} / a_{\beta_1 i})^2$
and $(a_{\beta_2 k} / a_{\beta_2 i})^2$.
We have three cases to consider:
\begin{itemize}
\item When $(a_{\beta_1 k} / a_{\beta_1 i})^2 > 1$
and $(a_{\beta_2 k} / a_{\beta_2 i})^2 > 1$, the $X_{\gamma}$
are imaginary and there are no $\tilde{S}$ that satisfy the spin
normalization constraints.
Said differently, the ellipsoid [Eq.~(\ref{eq:special-u-spin-norm1})]
and cylinder [Eq.~(\ref{eq:special-u-spin-norm2})] do not intersect.
On $0 < \kappa < \pi/3$,
this is the case for the $n=0$ solutions
on all $u^* = \mathbb{Z} \pi/3$.

\item When $(a_{\beta_1 k} / a_{\beta_1 i})^2 < 1$
and $(a_{\beta_2 k} / a_{\beta_2 i})^2 < 1$, $X_{\gamma}$
is real for all $v$, and Eqs.~(\ref{eq:spec-param1})-(\ref{eq:spec-param3})
parametrizes two ($\pm$) closed lines. Once projected on to the $i,j,k$ spin spheres,
one obtains lines that have a circular shape in the $\beta_1 \beta_2$
plane.
On $0 < \kappa < \pi/3$,
this is the case for the $n=1$ solutions
with $u^* = \mathbb{Z} 2 \pi/3$.
In detail, $(i,j) = (1,2)$ for
$u^* = 0$, $(1,3)$ for $u^* = 2\pi/3$, and
$(2,3)$ for $u^* = 4\pi/3$; $(\beta_1, \beta_2, \gamma) = (y,z,x)$
in all cases.

\item When $(a_{\beta_1 k} / a_{\beta_1 i})^2 > 1$
and $(a_{\beta_2 k} / a_{\beta_2 i})^2 < 1$, it is
better to eliminate $X_{\beta_2}$ from Eq.~(\ref{eq:special-u-spin-norm1}),
giving
\begin{eqnarray}
&&X_{\gamma}(v) =
\sqrt{1 - (a_{\beta_2 k} / a_{\beta_2 i})^2} \, \cos(v)\,, \\
&&X_{\beta_1}(v) =
\sqrt{\frac{(a_{\beta_2 i})^2 - (a_{\beta_2 k})^2}{(a_{\beta_2 i}/a_{\beta_1 i})^2 - (a_{\beta_2 k}/a_{\beta_1 k})^2}} \,
\sin(v)\,, \\
&&X_{\beta_2}(v) = \nonumber \\
&& \pm a_{\beta_1 k} 
\sqrt{\frac{1 - (a_{\beta_1 i} / a_{\beta_1 k})^2 \sin^2(v) - (a_{\beta_2 k} / a_{\beta_2 i})^2 \cos^2(v)}{(a_{\beta_1 k} / a_{\beta_2 k})^2 - (a_{\beta_1 i} / a_{\beta_2 i})^2}}\,. \nonumber \\
\end{eqnarray}
The above is again well-defined for all $v$, and
parametrizes two closed lines. Once projected on to the $i,j$ spheres,
we get lines circular in the $\gamma \beta_1$ plane. On the $k$ sphere,
we get a line squashed along
the $\gamma$ direction, $S^{\gamma}_k = 0$.
On $0 < \kappa < \pi/3$,
this is the case for the $n=1$ solutions
with $u^* = \pi/3 + \mathbb{Z} 2 \pi/3$.
In detail, $(i,j) = (2,3)$ for
$u^* = 0$, $(1,2)$ for $u^* = 2\pi/3$, and
$(1,3)$ for $u^* = 4\pi/3$;
$(\beta_1, \beta_2, \gamma) = (z, x, y)$
in all cases.
\end{itemize}

Lastly, let us comment on how the $i,j,k,\beta_1,\beta_2,\gamma$
are determined. Since from previous study we know
that $|a_{\beta i}|$ are continuous functions of $\kappa$
and that only at the special Ising and XXZ0 points
may additional $|a_{\beta i}|$ coincide or vanish,
it follows that
it is sufficient to determine the $i,j,k,\beta_1,\beta_2,\gamma$
for a given $u^* \in \mathbb{Z} \pi/3$
at one $\kappa \in \langle 0, \pi/3 \rangle$ to know these
indices across this whole range.
That is, because of discreteness and
continuity,
these indices may only change at $\kappa = \mathbb{Z} \pi/3$.

\subsection{Ground states for generic $u \neq \mathbb{Z}\pi/3$ and $0 < \kappa < \pi/3$}

For a generic $u$ and $\kappa$,
the null vectors of each $3\times3$ submatrix $M_{\beta}$ are
always given by Eq.~(\ref{eq:regular-null-vec}), and
$\tilde{S}$ can again be written as given
in Eq.~(\ref{eq:tilde-S-null-expr}).

The allowed values of $X_{\beta}$ are
determined by the intersection of 3 ellipsoids
that follow from the spin normalization constraints:
\begin{eqnarray}
\frac{X_x^2}{a_{xi}^2}+
\frac{X_y^2}{a_{yi}^2}+
\frac{X_z^2}{a_{zi}^2}=1\,, \quad i=1,2,3\,,
\label{eq:ellipsoids}
\end{eqnarray}
with semi-axis lengths $|a_{\beta i}|$ [Eq.~(\ref{eq:a-semi-axis-def})],
known to all be non-zero for $u \neq \mathbb{Z}\pi/3$.

One might expect Eqs.~(\ref{eq:ellipsoids}) to give eight solutions
for $X_{\beta}$ for each $\kappa, u, n$ given by
\begin{eqnarray}
&&\begin{pmatrix}
X_x^2 \\
X_y^2 \\
X_z^2
\end{pmatrix}
= \mathcal{A}^{-1} \cdot  
\begin{pmatrix}
1\\
1\\
1
\end{pmatrix}\,, \\
&&\mathcal{A}=
\begin{pmatrix}
\left( 1/a_{x1} \right)^2 & 
\left( 1/a_{y1} \right)^2 & 
\left( 1/a_{z1} \right)^2 \\
\left( 1/a_{x2} \right)^2 & 
\left( 1/a_{y2} \right)^2 & 
\left( 1/a_{z2} \right)^2 \\
\left( 1/a_{x3} \right)^2 & 
\left( 1/a_{y3} \right)^2 &
\left( 1/a_{z3} \right)^2
\end{pmatrix}\,.
\end{eqnarray}
However, the matrix $\mathcal{A}$ actually has a vanishing
determinant, signifying that one of the three constraints
in Eq.~(\ref{eq:ellipsoids}) is linearly dependent on the
other two.

In the previous section, we have already seen this
linear dependence make the $i,j$ constraints of Eq.~(\ref{eq:special-u-spin-norm1})
identical. Although we are presently unable to give an analytic proof of the
statement $\det \mathcal{A} =0$ for generic $u$,
we have verified it numerically for the full spectrum
of possible values of
$\kappa, u, n$ and it holds for all cases.

Given that only two of the constraints in Eq.~(\ref{eq:ellipsoids}) are linearly independent, there exist one-parameter families of solutions
for $X_{\beta}$.
We will parametrize these families with a continuous parameter $v$.
Combined with the fact that $u$ is also a
continuous parameter,
this means that for all sets of generic exchange parameters $\{J\}$
[Eq.~(\ref{eq:compactkappadef}) with $\kappa \neq \mathbb{Z}\pi/3$],
the set of ground states has two
continuous free parameters, $u$ and $v$.

For two ellipsoids with semi-axis lengths
$(|a_{x1}|, |a_{y1}|, |a_{z1}|)$ and
$(|a_{x2}|, |a_{y2}|, |a_{z2}|)$ (both having
principal semi-axes aligned along the coordinate directions $x,y,z$)
to intersect, each ellipsoid must have at least one axis
whose length is longer than the corresponding
axis of the other ellipsoid (i.e., we cannot have
$|a_{x1}|>|a_{x2}|$, $|a_{y1}|>|a_{y2}|$, $|a_{z1}|>|a_{z2}|$).

As previously established, $|a_{\beta i}|$ are
piecewise continuous functions of $u$ and $\kappa$
that can (for generic $\kappa$)
intersect only at the special $u^* = \mathbb{Z}\pi/3$
points. Thus if we consider, for instance,
$u \in \langle 0, \pi/3 \rangle$ and
$\kappa \in \langle 0, \pi/3 \rangle$, then
for this whole range all $|a_{\beta i}|$ differ
and, moreover, the corresponding ellipsoids must
intersect, or not, in the same way. When one considers
the $n=0$ solution, one must also pay attention to
the divergences of $|a_{\beta i}|$ that
happen at $u_{\rm div} = \pi/6 + \mathbb{Z}\pi/3$.

After analysing the various cases for $0 < \kappa < \pi/3$,
one finds that the $n=0$ ellipsoids
with semi-axes $|a_{\beta i}(u, n=0, \kappa)|$
never intersect. This agrees with the
previous section where the $n=0$ case
also could not satisfy the spin normalization constraints
(the cylinder and ellipsoid did not intersect).

As for the $n=1$ case, on the interval $0 < u < \pi$
(not including the special $\mathbb{Z}\pi/3$),
one finds that
\begin{eqnarray}
|a_{x1}|<|a_{x2}|\,,\  |a_{y1}|>|a_{y2}|\,,\  |a_{z1}|>|a_{z2}|\,,
\label{eq:assump_a1}
\end{eqnarray}
holds,
whereas for $\pi < u < 2 \pi$, the reverse holds:
\begin{eqnarray}
|a_{x1}|>|a_{x2}|\,,\  |a_{y1}|<|a_{y2}|\,,\  |a_{z1}|<|a_{z2}|\,.
\label{eq:assump_a2}
\end{eqnarray}

In either case, one may solve
Eq.~(\ref{eq:ellipsoids}) with $i=1$ for $X_x$,
\begin{eqnarray}
X_x = \pm a_{x1} \sqrt{
1 - \left( \frac{X_y}{a_{y1}} \right)^2
-\left( \frac{X_z}{a_{z1}} \right)^2
}\,\,,
\end{eqnarray}
and then substitute
in to Eq.~(\ref{eq:ellipsoids}) with $i=2$, yielding
\begin{eqnarray}
&&\left( \frac{X_y}{b_{y,x}^{2,1}} \right)^2+
\left( \frac{X_z}{b_{z,x}^{2,1}} \right)^2=1\,, \\
&& b_{\beta,\gamma}^{i,j} = b_{\gamma,\beta}^{j,i} =\sqrt{
\frac{(a_{\gamma j})^2-(a_{\gamma i})^2}
{
\left( a_{\gamma j} / a_{\beta j} \right)^2
-
\left( a_{\gamma i} / a_{\beta i} \right)^2
}
}\,\,.
\end{eqnarray}
Note how the inequalities (\ref{eq:assump_a1})
and (\ref{eq:assump_a2})
ensure that $b^{2,1}_{y,x}>0$
and $b^{2,1}_{z,x}>0$.

Altogether, 
the solutions of the three ellipsoid equations can be parametrized as:
\begin{eqnarray}
&&\hspace{-10pt}X_x(u,v,\kappa)= \nonumber \\
&&\hspace{-10pt} \pm a_{x1}(u,n=1,\kappa)
\sqrt{
1-
\left(
\frac{b^{2,1}_{y,x}}
{a_{y1}}
\right)^2 \cos^2(v)
-
\left(
\frac{b^{2,1}_{z,x}}
{a_{z1}}
\right)^2\sin^2(v)
}\,, \nonumber \\[-8pt]  
\label{eq:Xpmsigns} \\
&&\hspace{-10pt} X_y(u,v,\kappa)=
+ b^{2,1}_{y,x}(u,n=1,\kappa) \cos(v)\,, \quad 
\label{eq:Ypmsign} \\
&&\hspace{-10pt} X_z(u,v,\kappa)=
+ b^{2,1}_{z,x}(u,n=1,\kappa) \sin(v)\,. \quad 
\label{eq:Zpmsign}
\end{eqnarray}

If we consider the $\tilde{S}$ that the
above solutions for $X_{\beta}$,
when combined with
Eqs.~(\ref{eq:exp_val}), (\ref{eq:Stilde}),
and (\ref{eq:tilde-S-null-expr}),
give, then one finds that for a fixed sign of $X_x$
these solutions are discontinuous at $u = \mathbb{Z} 2 \pi / 3$.
The cause is the fact that $X_x$ always has the same sign,
whereas $a_{xi}$, and therefore $S^x_i$, change their sign
at $u = \mathbb{Z} 2 \pi / 3$.

If one tries to find a sign convention that preserves continuity,
then one finds that the six patches of length $2\pi/3$ that
Eqs.~(\ref{eq:Xpmsigns})-(\ref{eq:Zpmsign}) yield
combine in to one periodic
patch of length $4\pi$.

In addition to the above, if one considers $\tilde{S}$ for
a fixed $v$ as a function of $u$, the $+$ sign convention
in Eq.~(\ref{eq:Ypmsign}) again gives discontinuities,
but this time at $u=\pi/3+\mathbb{Z}2\pi/3$ and in
the $S^y_i$ components of the spins. A simple sign
change eliminates these.

In summary, the $X_{\beta}$,
now named $X,Y,Z$,
that give a continuous parameterisation
of the ground state manifold are
\begin{eqnarray}
&&\hspace{-16pt} X(u,v,\kappa)=
{\rm sgn}\left[\sin(3u/2)\right] a_{x1}(u,\kappa) \times \nonumber \\
&&\hspace{-16pt}  \qquad 
\sqrt{
1-
\left(
\frac{b^{2,1}_{y,x}}
{a_{y1}}
\right)^2 \cos^2(v)
-
\left(
\frac{b^{2,1}_{z,x}}
{a_{z1}}
\right)^2\sin^2(v)
}\,, \\
&&\hspace{-16pt} Y(u,v,\kappa)=
{\rm sgn}\left[\cos(3u/2)\right] b^{2,1}_{y,x}(u,\kappa) \cos(v)\,, \\
&&\hspace{-16pt} Z(u,v,\kappa)=
+ b^{2,1}_{z,x}(u,\kappa) \sin(v)\,,
\end{eqnarray}
where $u$ spans $[0, 4\pi]$ and $v$ spans $[0, 2\pi]$.
By substituting in to Eq.~(\ref{eq:tilde-S-null-expr}),
one obtains the desired values of all
possible spins, and through Eq.~(\ref{eq:exp_val})
product states, that are in the ground state:
\begin{eqnarray}
&&{\bf S}_1(u, v, \kappa) =
\frac{1}{2}
\begin{pmatrix}
\displaystyle \frac{X(u, v, \kappa)}{a_{x1}(u, \kappa)} \\
\displaystyle \frac{Y(u, v, \kappa)}{a_{y1}(u, \kappa)} \\
\displaystyle \frac{Z(u, v, \kappa)}{a_{z1}(u, \kappa)}
\end{pmatrix}\,, \label{eq:final-param-of-M-1} \\
&&{\bf S}_2(u, v, \kappa) =
\frac{1}{2}
\begin{pmatrix}
\displaystyle \frac{X(u, v, \kappa)}{a_{x2}(u, \kappa)} \\
\displaystyle \frac{Y(u, v, \kappa)}{a_{y2}(u, \kappa)} \\
\displaystyle \frac{Z(u, v, \kappa)}{a_{z2}(u, \kappa)}
\end{pmatrix}\,, \\
&&{\bf S}_3(u, v, \kappa) =
\frac{1}{2}
\begin{pmatrix}
\displaystyle \frac{X(u, v, \kappa)}{a_{x3}(u, \kappa)} \\
\displaystyle \frac{Y(u, v, \kappa)}{a_{y3}(u, \kappa)} \\
\displaystyle \frac{Z(u, v, \kappa)}{a_{z3}(u, \kappa)}
\end{pmatrix}\,. \label{eq:final-param-of-M-3}
\end{eqnarray}

The above expressions are valid even at the special $u$ points
if one takes a careful limiting procedure
$u \to u^* \in \mathbb{Z}\pi/3$.
What one obtains through such a limiting procedure
agrees with the solutions derived in the previous
section explicitly.
This confirms that the above solutions
exhaust the manifold of all product state ground states
for $\kappa \neq \mathbb{Z}\pi/3$.

In the Supplementary Material~\cite{supplemental},
the reader may also find
animations of the parameterisation
from Eqs.~(\ref{eq:final-param-of-M-1})-(\ref{eq:final-param-of-M-3})
for $\kappa = 0.1, \pi/6, 0.9$.
"In the animations, two rows are drawn with three spin spheres and one $\mathcal{M}$ torus in each. In the upper row, lines of constant $u$ are drawn whose coloring indicates how $v$ is varied along them. In the lower row, $v$ is held constant and $u$ is varied."

\section{Topology of the manifold of exact product ground states on a single triangle}
\label{app:topology}

In the previous section, we have found explicit
parameterisations of all possible product states [Eq.~(\ref{eq:product-wf})]
that are grounds states of a single triangle.
That is, at the points in
the $\{J\}$ parameter space where the ground
state is six-fold degenerate [Eq.~(\ref{eq:kappadef})],
we have found that this
six-dimensional subspace of ground states
includes a two-dimensional submanifold
of separable states.
Here we shortly discuss the topology of this
separable ground state manifold $\mathcal{M}$.


The manifold $\mathcal{M}$ can be looked at
as a submanifold
embedded in the quantum
space of ground state rays
$\mathbb{C} \mathbf{P}^5 = \mathbb{C}^6/{\sim} = \mathcal{S}^{11} / {\rm U}(1)$,
but also as a submanifold embedded in
the classical configuration space
$\mathcal{S}^2 \times \mathcal{S}^2 \times \mathcal{S}^2$.
The latter is just the statement that the product
states are in correspondence with a state of classical vector spins, with components corresponding to the expectation values of the spin operators.

Although a small portion of $\mathbb{C} \mathbf{P}^5$,
superpositions of separable states contained in $\mathcal{M}$
are in fact sufficient to describe all ground states
on a single triangle.
For the Ising and XXZ0 points, this
can be verified
explicitly, whereas for generic $\kappa$
we have verified this statement numerically
for all $\kappa$:
by taking six or more random points
of $\mathcal{M}$ and calculating their
corresponding quantum state vectors,
one finds that the rank of these
vectors is always six.

As shown in Fig.~(\ref{fig:special_point_topology}),
the $\mathcal{M}$ of Ising and XXZ0 points
are made of a series of spheres
that touch neighbors in just
such a way that together they
resemble a wreath.
In general, upon varying $\kappa$,
the instabilities that these points of contact
represent
can be resolved by either combining
or disconnecting these spheres.
Because of
the spin permutation
$({\bf S}_1, {\bf S}_2, {\bf S}_3) \mapsto ({\bf S}_{\pi(1)}, {\bf S}_{\pi(2)}, {\bf S}_{\pi(3)})$
and inversion ${\bf S}_i \mapsto - {\bf S}_i$
symmetries, what happens at one junction
must happen at all the other junctions.

Since there are no special $\kappa$ points
besides these points,
and there is no continuous way one can
transform six disjoint spheres
in to two touching spheres,
it seems intuitively clear that the spheres
must combine into one torus.
And indeed, in our exact solution
we have found that
(for generic $\kappa$) $\mathcal{M}$
can be smoothly parametrized with
two periodic parameters $u \in [0,4\pi]$
and $v \in [0,2\pi]$.
Thus, $\mathcal{M}$ is topologically a torus.
We have also verified this numerically
(see next section for details).

Although both the Ising and XXZ0 points
arise from a thinning out of the torus $\mathcal{M}$ at certain rings,
for the Ising points the thinning happens
at constant $u$, while for the XXZ0 points
the thinning happens for constant $v$.
This should be clearly visible in the
animations of $\mathcal{M}$, available
in the Supplementary Material~\cite{supplemental}.

Let us also comment on
the mapping
${\bf S}_1\colon \mathcal{M} \to \mathcal{S}^2$.
For generic $\kappa$, the Brouwer degree
of this mapping is
${\rm deg} \, {\bf S}_1 = \frac{1}{4 \pi} \int_{\mathcal{M}} d\Omega_1 = 2$,
where $d\Omega_1$ is the spherical angle
two-form.
Because of the parameterisation
permutation symmetry [Eqs.~(\ref{eq:Jcoef-perm1})-(\ref{eq:Jcoef-perm2})]
and smoothness in $\kappa$,
it is sufficient to
determine this numerically at one point
to conclude that it
holds for all $\kappa \neq \mathbb{Z}\pi/3$.


One can also consider the ${\rm U}(1)$-bundle
associated with the Hilbert space modulo
global phases. As usual for spin-half particles,
the curvature of the connection
of this bundle is
$F = \frac{1}{2} (d\Omega_1 + d\Omega_2 + d\Omega_3)$.
It is interesting to note how once one
considers the analogous ${\rm U}(1)$-bundle
of $\mathcal{M}$, the
first Chern number is
$c_1 = \frac{1}{2 \pi} \int_{\mathcal{M}} F = {\rm deg} \, {\bf S}_1 + {\rm deg} \, {\bf S}_2 + {\rm deg} \, {\bf S}_3 = 6$,
implying that the ${\rm U}(1)$-bundle
of the torus $\mathcal{M}$
is non-trivial.

\subsection{Finding the topology of closed 2D manifolds numerically}

The numerical procedure of determining
the topology of 2D closed manifolds relies
on the classification theorem that
the orientability and genus uniquely
determine the topology of every closed
(compact without a boundary)
2D manifold~\cite{Edelsbrunner-topo}.

In 2D, the possible closed manifolds
are the sphere (orientable, genus $g=0$),
connected sums of tori (orientable, $g \geq 1$),
and connected sums of real projective planes
(non-orientable, non-orientable genus $k \geq 1$).

Since numerically, in general, the only thing that
one can do is determine whether a point is in the
manifold of interest, the question is how to
determine from a collection of points (a mesh)
the orientability and genus.

This can be done if we can find a proper
triangulation of this mesh.
By a proper triangulation
we mean a triangulation
whose edges all border two, and only two,
distinct triangles.

If we can consistently attribute an orientation
to all triangles of the triangulation,
then the manifold is orientable.
Let us recall that an orientation of a triangle
is given by attributing directions
to all the triangle edges in a circular way.
The orientations of adjacent triangles are
consistent if the directions of the shared edge
are opposite.
Moreover,
from the number of vertices $V$, edges $E$, and
triangles $T$, we can calculate the
Euler characteristic $\chi = V - E + T$.
The Euler characteristic is related to
the genus through $\chi = 2 - 2g$,
and to the non-orientable genus
through $\chi = 2 - k$.
Thus finding the topology of $\mathcal{M}$
amounts to finding a proper triangulation
of $\mathcal{M}$.

Although there are various procedures
for finding the triangulation of a
collection of points,
we have implemented the following
simple procedure (see the supplemented file
{\ttfamily triangulation\_finder.py} for a Python implementation~\cite{supplemental}):
\begin{itemize}
\item Starting from a random point $i$
of the manifold mesh,
we first find the closest $\sim 32$ points
of the mesh.
\item Then we find the best fitting plane
of these closest points
through a singular-value decomposition,
and project all the $\sim 32$
closest point on to this plane.
\item Next, we find the Delaunay triangulation
of this 2D sample of points and
add the edges and triangles
of only those points that are adjacent to $i$
to the manifold triangulation.
\item Lastly, we add these points adjacent to
$i$ to a queue, and repeat the above
three steps for
all points in the queue.
The only difference is that now we have
to carry out a constrained Delaunay triangulation
to ensure that we respect the edges
of the previous local-plane triangulations.
\end{itemize}
A demonstration of our numerical
routine,
together with an application to
an already computed mesh of
the ground state manifold $\mathcal{M}$,
is given in
{\ttfamily triangulation\_demonstration\_and\_verification.ipynb}.

Our routine detects various cusps or
irregularities in the manifold or
mesh by calculating the standard deviations
orthogonal to the best fitting plane.
If they are large, a warning is given.

In addition, it is also a good idea to
use a mesh that is dense and uniform.
Uniformity can be achieved by relaxing
the mesh under the influence of short-ranged
repulsive forces.

When the {\ttfamily triangulation\_finder.py}
routine is applied to a sufficiently
dense mesh of the ground state manifold $\mathcal{M}$,
for $\kappa$
not too close to the Ising or XXZ0 points,
one finds that $\mathcal{M}$ is orientable and
has genus $1$. Thus $\mathcal{M}$ is a torus.

\bibliography{refs.bib}
\end{document}